\documentclass[twocolumn,showpacs,preprintnumbers,amsmath,amssymb,prl,superscriptaddress]{revtex4-1}
\pdfoutput=1
\usepackage{graphicx}
\usepackage{dcolumn}
\usepackage{bbold}
\usepackage{color}

\usepackage[low-sup]{subdepth}
\usepackage{multirow}
\usepackage{amsfonts}
\usepackage[english]{babel}
\usepackage[T1]{fontenc}
\usepackage{times}
\usepackage[colorlinks,bookmarks=true,citecolor=blue,linkcolor=red,urlcolor=blue]{hyperref}
\usepackage[tight, FIGTOPCAP, hang, raggedright, nooneline]{subfigure}

\usepackage{verbatim}
\usepackage{bm}
\usepackage{mathrsfs}

\newcommand{\ket}[1]{\left|#1\right>}
\newcommand{\bra}[1]{\left<#1\right|}

\newcommand{\para}[1]{\left(#1\right)}
\newcommand{\abs}[1]{\left|#1\right|}

\newcommand{\COMMENT}[1]{}

\def\beq{\begin{equation}}
\def\eeq{\end{equation}}
\def\barray{\begin{eqnarray}}
\def\earray{\end{eqnarray}}

\begin{document}
\title{ Entanglement distance between quantum states and its implications for density-matrix-renormalization-group study of degenerate ground-states}
\author{Mohammad-Sadegh Vaezi}
\affiliation{Department of Physics, Washington University, St. Louis, MO 63160, USA}
\author{Abolhassan Vaezi}
\email{vaezi@stanford.edu}
\affiliation{Department of Physics, Stanford University, Stanford, CA 94305, USA}

\begin{abstract} 
We study the concept of entanglement distance between two quantum states which quantifies the amount of information shared between their reduced density matrices (RDMs).  Using analytical arguments combined with density-matrix-renormalization-group (DMRG) and exact diagonalization (ED) calculations, we show that for gapless systems the entanglement distance has power law dependence on the energy separation and subsystem size with $\alpha_E$ and $\alpha_{\ell}$ exponents, respectively. Using conformal field theory (CFT) we find $\alpha_E = 2$ and $\alpha_{\ell} = 4$ for Abelian theories with $c=1$ such as free fermions. For non-Abelian CFTs $\alpha_E = 0$ , and $\alpha_{\ell}$ is twice the conformal dimension of the thermal primary fields. For instance for $Z_3$ parafermion CFT $\alpha_E = 1$ and $\alpha_{\ell} = 4/5$. For gapped 1+1D fermion systems, we show that the entanglement distance divides the low energy excitations into two branches with different values of $\alpha_E$ and $\alpha_{\ell}$. These two branches are related to momentum transfers near zero and $\pi$. We also demonstrate that the entanglement distance reaches its maximum for degenerate states related through nonlocal operators such as Wilson loops. For example, degenerate ground-states (GSs) of 2+1 D topological states have maximum entanglement distance. On the contrary, degenerate GSs related through confined anyon excitations such as genons have minimum entanglement distance. Various implications of this concept for quantum simulations are discussed. Finally, based on the ideas developed we discuss the computational complexity of DMRG algorithms that are capable of finding all degenerate GSs. 
\end{abstract}

\maketitle

\section{Introduction}
Entanglement-based quantum simulations such as DMRG, tensor product states (TPS) and multi-scale entanglement renormalization ansatz (MERA) have revolutionized our understanding of low dimensional quantum systems \cite{White1992a,Ostlund1995a,Vidal2007a,Cirac2008a,Schollwock2011a,Stoudenmire_2012a,Vidal2004a,White_Feiguin2004a,Vidal2004b,White1996a,McCulloch2008a,White2005a}. These approaches are built on the fact that the GS of local Hamiltonians has a significantly lower complexity measured in units of entanglement entropy (EE) than a generic excited state allowing a more efficient data compression \cite{Eisert2010a}. On the other hand, the entanglement related quantities themselves have become an essential tool in the characterization of the GSs and low energy excitations  \cite{Holzhey1994a,Calabrese2004a,Kitaev2004a,Levin2004a,Li_Haldane2008a,Pollmann2010a,Sterdyniak2011a,Zhang_Grover2012a,Jiang_2008a,Jiang2012a,Depenbrock2012a,Yan_Huse2011a,Ryu2006a,Swingle2012a,Zaletel_2013a,Liu_Vaezi2015a}. The standard approach in most quantum simulation algorithms targets a single energy eigenstate, usually a GS (among possibly several ones), e.g., through constructing the projection (truncation) matrices using the RDM associated with that particular GS. In other words, the standard approach is non-ergodic and as a result the information about other potentially degenerate GSs or excited states is partially or completely lost. This is the main reason why single-state targeting DMRG cannot necessarily obtain all degenerate GSs. Nevertheless, having access to all GSs is crucial in studying topological order e.g., to obtain modular matrices, or the fusion rules and braid statistics of anyons~\cite{Wen1990a,Nayak2008a}. 

Here, we systematically address this problem and introduce two multiple-states targeting DMRG algorithms capable of accessing all degenerate GSs. In particular we explore two metrics for measuring what we will refer to as {\em entanglement distance} between two energy eigenstates which quantifies the amount of information encoded in the targeted state about other states. We use DMRG, ED, analytical approaches and in the case of non-interacting fermions exact results for fairly large system sizes to study the behavior of the entanglement distance in various systems. We show that the entanglement distance exhibits distinct behaviors in each of the following classes: (a) gapless, (b) trivial gapped, and (c) topological gapped states. Furthermore, we show that it provides an alternative way of measuring conformal dimensions in 1+1D CFTs such that the finite size effect is less significant compared with other methods.

\begin{figure}
\centerline{\includegraphics[width  =1.0\linewidth] {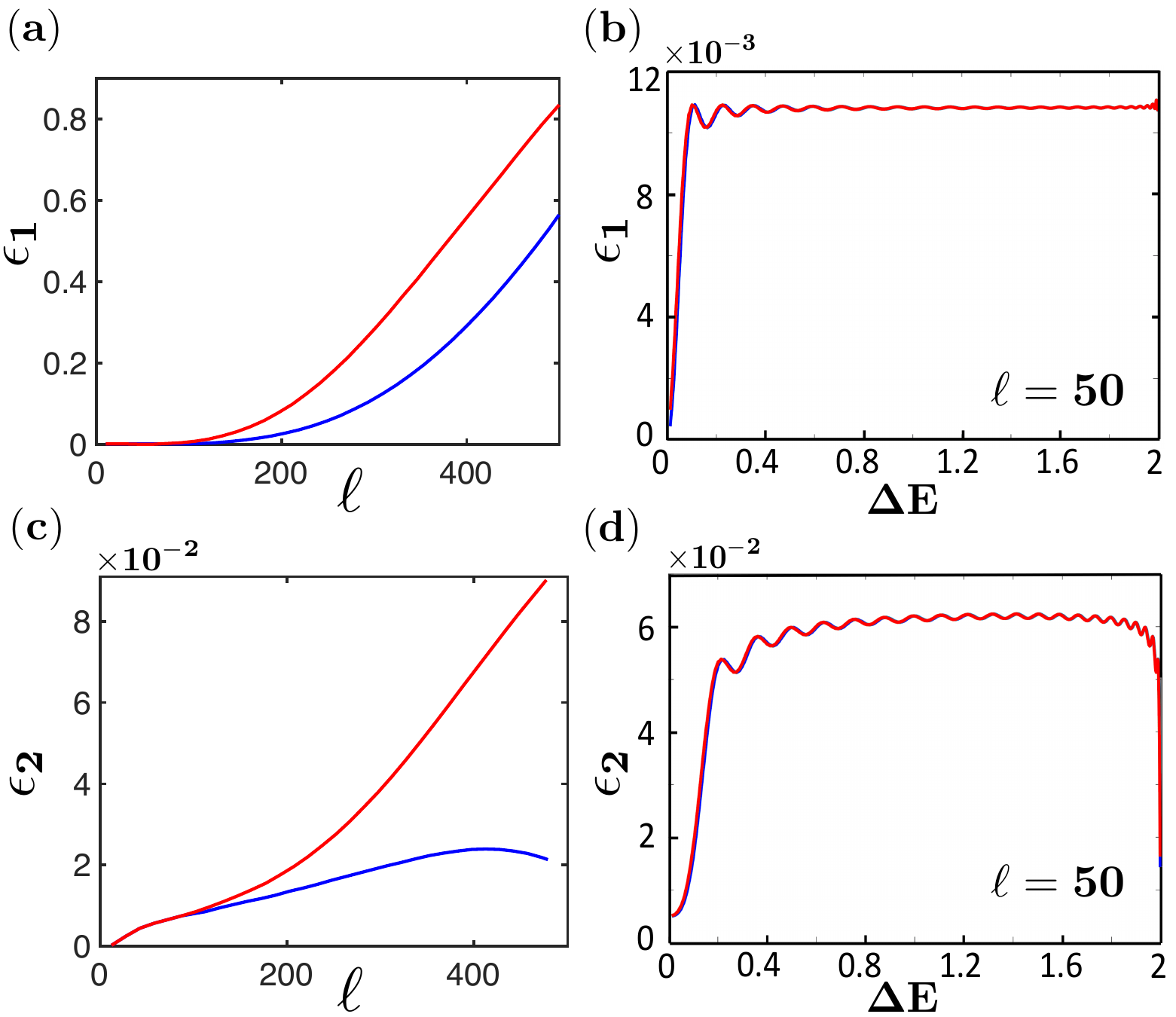}}
\caption{\label{fig1}  The two measures of entanglement distance between the GS and excited states obtained by creating an electron hole excitation $(c_{k_F+\Delta k}^\dag c_{k_F})$ in a gapless 1+1D free fermions with nearest neighbor hopping $t_1 = 1$ at half-filling, and $\mathcal{N} = 1000$. The blue (orange) color represents excitations with $\abs{\Delta k}<\pi/2$ ($>\pi/2$) momentum transfer. Insets are log-log plots showing the power law behavior for small $\ell$ and $\Delta E$. (a) $\epsilon_1$ for the two lowest excited states vs $\ell$ .b) $\epsilon_1$ for all electron-hole excited states ($\ell = 50$). There are oscillations of period $\mathcal{N}/\ell$ around the saturation point. (b) and (c) similar quantities for $\epsilon_2^{\chi}$ with $\chi = 2^6$. Both measures for entanglement distance grow monotonically with $\ell$. They also start growing for small $\Delta E$ and then saturate and oscillate around the saturation point. The power law growth of entanglement distance for small $\ell$ and $\Delta E$ suggests: $\alpha_E = 1.9$ (1.6), $\alpha_{\ell} = 3.9$ (3.9), $\beta_{\ell}(2^6) = 0.86$~(0.98) for the blue (orange) branch, close to our theoretical predictions.}
\end{figure}

\begin{figure}
\centerline{\includegraphics[width  =1.0\linewidth] {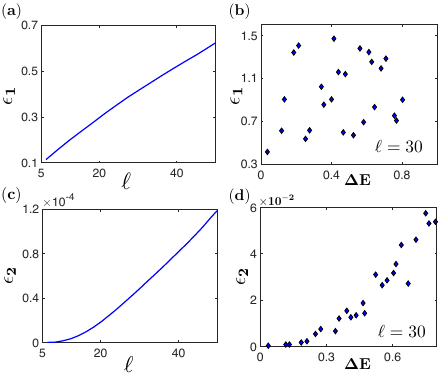}}
\caption{\label{fig2}  DMRG results with $\chi = 20$ for the entanglement distance metrics between the GS and excited states of a critical $Z_3$ parafermion chain (equivalent to a 3-state Potts model) of length $\mathcal{N} = 100$. These results suggest that $\alpha_E$ is approximately vanishing and $\alpha_{\ell} = 0.82$ indicating the scaling dimension of the thermal operator must be around $0.41$, both consistent with our theoretical predictions. Also,  we find that $\beta_E(\chi) = \chi/20 + 1.3$  ($\ell = 30$) with a high accuracy (see Appendix B for more results) and $\beta_{\ell}(\chi)$ has a weaker $\chi$ dependance.}
\end{figure}

\section{Entanglement distance} Here we consider two distinct entanglement based metrics for gauging the distance between a pair of quantum states. Consider $\ket{\Psi}_a$ and $\ket{\Psi}_b$ eigenstates of a (local) Hamiltonian $\mathcal{H}$ defined on a connected manifold $\mathcal{M}$ which is bipartitioned into $L$ and $R$ subsystems.  The (left) RDM associated with state $a$ is given by: $\rho_L^{(a)} = {\rm tr}_R\para{\ket{\Psi}_a\bra{\Psi}_a}$. 
We define
\begin{eqnarray}
\epsilon_{1}\para{a,b} \equiv     {\mathrm{tr~}\para{\rho_L^{(a)}-\rho_L^{(b)}}^2 }/{ \mathrm{tr~}{\rho_L^{({\mathbb{1}})}}^2},~~\label{EPS1}
\end{eqnarray}
as the first measure of entanglement distance, where $\rho_L^{({\mathbb{1}})}$ denotes the GS's RDM.  The second measure is inspired by DMRG and TPS quantum simulation algorithms. Let us consider RDM $\rho_L^{\para{a}}$ with dimension $\mathcal{D}_L$ and we denote its eigenvectors and eigenvalues by $\ket{v}_{l,\para{a}}$ and $\lambda_{l,\para{a}}$ respectively. 
The set of $\chi_a$ (the so-called bond dimension in TPS) dominant eigenvalues of  $\rho_L^{\para{a}}$ form a matrix $T_L^{\para{a}}$ whose dimension is $\mathcal{D}_L\times \chi_a$ and can be used for truncating operators and states~(see
Appendix A for more details on the DMRG method). It acts on a generic $\mathcal{D}_L \times \mathcal{D}_L$ dimensional operator $\mathcal{O}_L$ with support on region $L$ and yields $\overline{\mathcal{O}}_L = {T_L^{\para{a}}}^\dag \mathcal{O}_L T_L^{\para{a}}$ with a lower dimension, $\chi_a \times \chi_a$. Demanding the correlation functions of $\mathcal{O}_L$ to remain nearly invariant after projection imposes certain constraints on the lower bound of the bond dimension, $\chi_a$. It is generally believed that $\chi_{a,\rm min} \sim e^{S_L}$, where $S_L$ is the EE between the two subsystems. Furthermore, by construction, $T_L^{\para{a}}$ minimizes the following cost function known as truncation error: $\epsilon_{2}^{\para{\chi_a}}\para{a,a} \equiv 1 - \mathrm{tr}\para{T_L^{\dag}\rho_L^{\para{a}}T_L} $. Since, $\rho_L^{\para{b}}$ played no role in defining $T_L^{\dag}\para{a}$ one may wonder how the following quantity behaves~\cite{Comment4}:
\beq
\epsilon_{2}^{\para{\chi_a}}\para{a,b} \equiv 1 - \mathrm{tr}\para{{T_L^{\para{a}}}^{\dag}\rho_L^{\para{b}}T_L^{\para{a}}} .\label{EPS2}
\eeq
Indeed in general it is not at all clear how efficient $T_L^{(a)}$ is in preserving information stored in $\rho_L^{\para{b}}$ (e.g., in reproducing correlation functions). It is quite possible that it may discard most of dominant eigenvectors
of $\rho_L^{\para{b}}$ and instead retain the subdominant ones.  We will see that for gapless systems the low energy excitations exhibit $\epsilon_1 \propto \Delta E^{\alpha_E}\ell^{\alpha_{\ell}}$, and $\epsilon_2^{(\chi)}\propto \Delta E^{\beta_E(\chi)}\ell^{\beta_{\ell}(\chi)}$ power law behaviors where $\ell$ is the left subsystem size and $\Delta E$ the excitation energy.

\subsection{A.  Entanglement distance in 1 + 1 D CFTs}
We consider conformal field theories in 1+1D systems of length $\mathcal{N}$. The excited states can be obtained by acting primary or descendant fields on the GS (vacuum). The EE of the GS and excited states are obtained in Refs. \cite{Holzhey1994a,Calabrese2004a,Alcaraz2011a,Berganza2011a,Taddia2016a,Astaneh2013a}. Here, we are interested in finding the entanglement distance between the GS and an excited state associated with $\Upsilon$ primary field with conformal weights $h$ and $\bar{h}$ . To this end, we closely follow the approach and notations of Ref. \cite{Alcaraz2011a}.

The excited state can be related to GS as : $|\Upsilon \rangle = \lim_{z, \bar{z} \rightarrow -i\infty}  \Upsilon(z, \bar{z}) \, | 0 \rangle$. The wave function of this state has the following path integral representation:
$\Psi_{XY}(\Upsilon) \propto \int {\cal D} \phi \;  \Upsilon[ \phi(z_\infty)]    \; e^{ - S(\phi)} $ where $X$ ($Y$) denotes the coordinates on the left (right) region, and $\phi$ is the local dynamical field whose Euclidean action is $S(\phi)$. Similarly, the RDM associated with subsystem $L$ is: $\rho_L^{\Upsilon}\para{X X' }  \propto \int {\cal D} Y \;  \Psi_{XY}(\Upsilon) \, 
\Psi_{YX'}^*(\Upsilon) $. After normalization:
\beq
\rho_L^{\Upsilon}\para{X X' } =   \frac{  \int {\cal D} \phi  \;  \Upsilon[ \phi(z_\infty)]  \,  \Upsilon^*[ \phi(z'_\infty)]  \; e^{ - S(\phi)} }{ Z(1)  \langle \Upsilon(z_\infty) \, \Upsilon^\dagger( z'_\infty) \rangle}. 
  \label{Rho1}
\eeq
Now we need to compute $\epsilon_{1}\para{\Upsilon_a,\Upsilon_b}$ which requires computing $M_{ab} \equiv \mathrm{tr~}\para{\rho_L^{\Upsilon_a}\rho_L^{\Upsilon_b}}$ first. Similar to the well-established procedure of evaluating EE of CFT states this quantity can be transformed into a path integral. The resulting path integral is defined on a manifold which is formed of two cylinders on the right subsystem and a single two-sheeted Riemann surfaces which is identical to a single cylinder whose radius is twice larger than cylinders on the right side. The two submanifolds are glued at the boundaries. A conformal transformation can be applied to push the boundaries between $L$ and $R$ subsytems to infinity after which we are left with $L$ subsystem. Evaluating the path integral on the resulting manifold, we arrive at the following relation:

\begin{widetext}
\barray
&&F^{(2)}_{\Upsilon_a\Upsilon_a}  \equiv \frac{\mathrm{tr}~ \rho_L^{\Upsilon_a} \rho_L^{\Upsilon_b} }{\mathrm{tr}~\rho_L^{\mathbb{1}} \rho_L^{\mathbb{1}}} =
 \frac{ \langle   \Upsilon_a(0) \, \Upsilon_a^\dagger\para{\pi x} \, \Upsilon_b^\dagger\para{\pi } \Upsilon_b^\dagger\para{\pi (1 + x)} \rangle_{\rm cy}}
{  
  2^{ 2 ( h_a+h_b + \bar{h}_a + \bar{h}_b)}\langle \Upsilon_a(0) \, \Upsilon_a^\dagger( 2 \pi x) \rangle_{\rm cy}\langle \Upsilon_b(0) \, \Upsilon_b^\dagger( 2 \pi x) \rangle_{\rm cy}},  \nonumber  \label{F2aa}
\earray

\end{widetext}

where $x = \frac{\ell}{\mathcal{N}}$. The two-point correlation function of primary and descendant fields on cylinder is:  
$\langle \Upsilon_j(w_1,\bar{w}_2)\Upsilon_j^\dag(w_2,\bar{w}_2) \rangle \propto {\para{2\sin(\frac{w_1-w_2}{2})}^{-2h_j}\para{2\sin(\frac{\bar{w}_1-\bar{w}_2}{2})}^{-2\bar{h}_j}}$.
Plugging this relation into Eq. \eqref{F2aa} we obtain:
\barray
F^{(2)}_{\mathbb{1}\Upsilon_b}  = \para{\cos{\frac{\pi x}{2}}}^{2\para{h_b+\bar{h}_b}} ~\sim~1 - \frac{h_b+\bar{h}_b}{4}\para{\pi x}^2+O(x^4).~~\label{F21b} \cr
&&
\earray
Similarly, $F^{(2)}_{\Upsilon_b\Upsilon_b}$ can be obtained using the Wick's theorem and $\Upsilon_b \times \Upsilon_b^\dag={\bf 1}+\Psi+\ldots$ OPE. According to Ref. \cite{Alcaraz2011a}, assuming $\Psi$ is the operator with the smallest scaling dimension $\Delta_\Psi$ and OPE coefficient $C^{\Psi}_{\Upsilon_b\Upsilon_b^\dag}$, in the $x<<1$ limit: 
\beq
 F^{(2)}_{\Upsilon_b\Upsilon_b} \sim 1-\frac{h_b+\bar{h}_b}{2}(\pi x)^2+C^{\Psi}_{\Upsilon_b\Upsilon_b^\dag}(x^{2\Delta_\Psi})+\cdots~.
\label{F2bb}
\eeq
Combining the above results, the first measure of entanglement distance becomes: 
\barray
 \epsilon_1\para{\Upsilon_b,\mathbb{1}} \sim C^{\Psi}_{\Upsilon_b\Upsilon_b^\dag}x^{2\Delta_\Psi} + c_2 \para{h_b+\bar{h}_b}^2 x^4 + \cdots ~,
\label{EPS3}
\earray
where $c_2$ is a constant. Recall that in CFT the excitation energy is proportional to $\frac{2\pi \para{{h}_b + \bar{h}_b}}{\mathcal{N}}$.
Thus, the above results suggest that for non-Abelian CFTs where $\Psi$ is a non-trivial primary field, $\epsilon_1\para{\Upsilon_b,\mathbb{1}} \sim x^{2\Delta_{\Psi}}$ and almost insensitive to $\Delta E$ to the lowest order of $x$ and $\Delta E$, while for Abelian CFTs where $\Upsilon_b \times \Upsilon_b^\dag={\bf 1}$, ~~  $\epsilon_1\para{\Upsilon_b,\mathbb{1}} \sim x^{4}\Delta E^2$ with higher order corrections. Using exact computations for non-interacting fermion systems as well as DMRG study of parafermion chains \cite{Fradkin_kadanoff1980a,Fendley2012a,Ortiz_Cobanera2011a,Vaezi_Kim2013a},  these two distinct behaviors can be verified (see Figs. \ref{fig1} and \ref{fig2}). For example, we find $\Delta_{\Psi} = \alpha_{\ell}/2= 0.41$ for the $Z_3$ parafermion chain which is extremely close to the conformal dimension of the thermal operator $(\Delta_{\epsilon} = 2/5)$.

\begin{figure}
\centerline{\includegraphics[width  =1.0\linewidth] {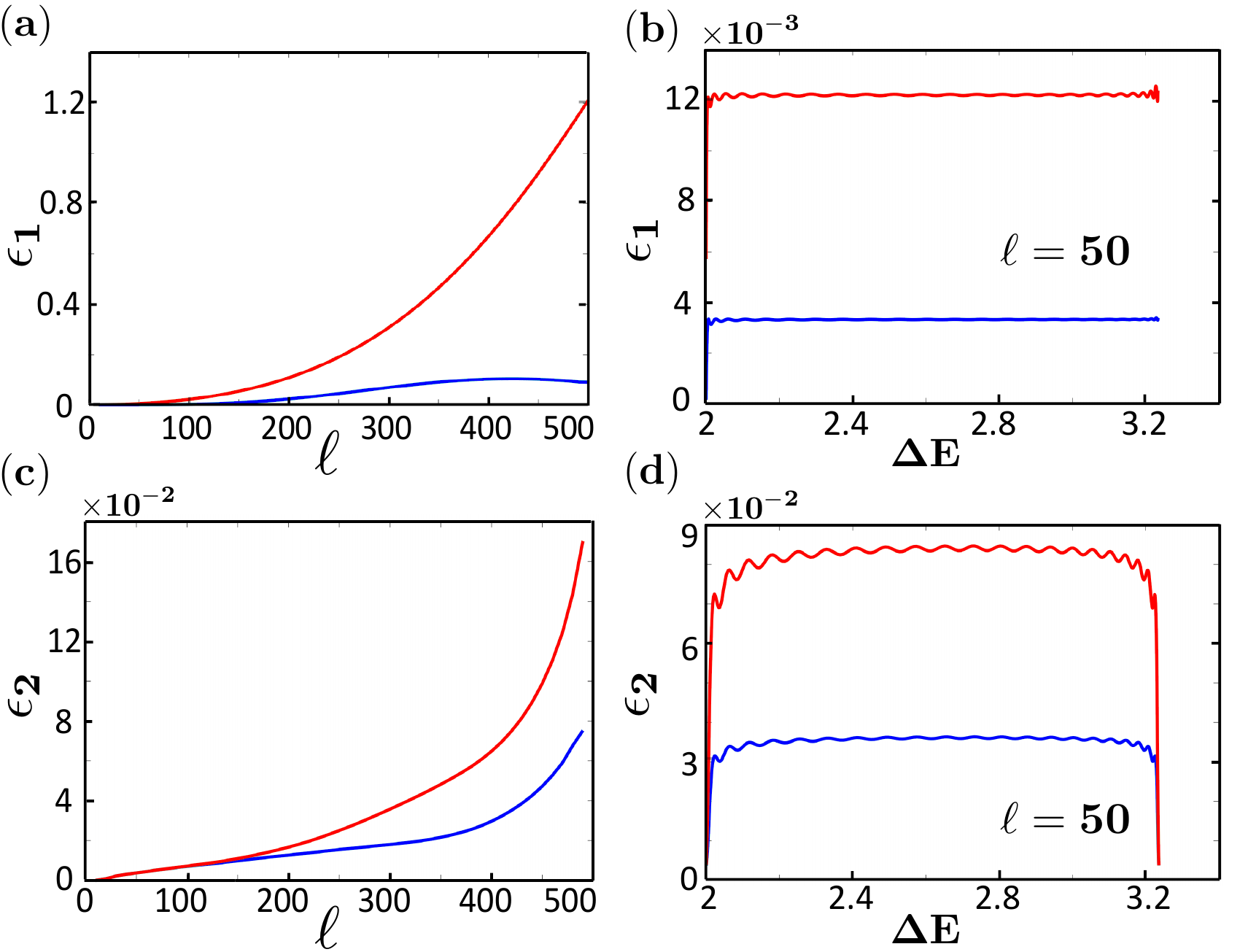}}
\caption{\label{fig3}  The same plot as Fig. \ref{fig1} for a gapped 1+1D free fermions with staggered chemical potential $\mu_n = (-1)^n$. The excited states are bifurcated into two branches distinguished by the momentum transfer. 
The power law behavior of the entanglement distance suggests: $\alpha_{\ell} = 3.8$~(2.2) for the blue (orange) branch. }
\end{figure}

\begin{figure}
\centerline{\includegraphics[width  =1.0\linewidth] {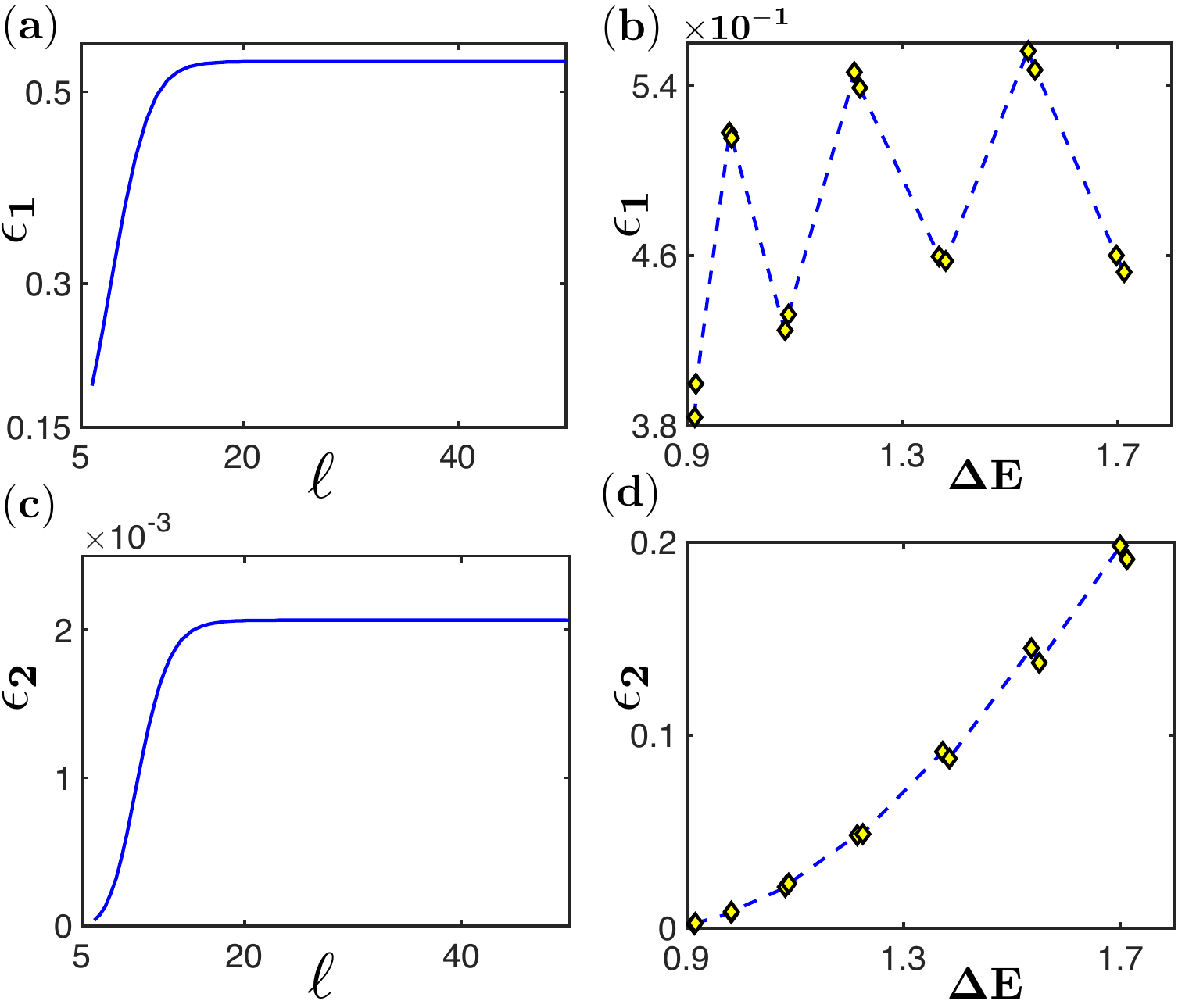}}
\caption{\label{fig4}  The same plot as Fig. \ref{fig2}, for a gapped $Z_3$ parafermion chain in the trivial phase. Both metrics of entanglement distance saturate when $\ell$ becomes comparable with the correlation length.}
\end{figure}

Finding an analytic expression for $\epsilon_2^{\chi_{\mathbb{1}}}\para{\mathbb{1},\Upsilon_b}$ is more challenging. Instead, we use DMRG to study the behavior of this quantity in the 1+1D gapless states of free fermions and $Z_3$ parafermion chain. (see Fig. \ref{fig1} and \ref{fig2}). Again, we observe a power law behavior as a function of energy separation as well as $\ell/\mathcal{N}$ but with different ($\chi$-dependent) exponents. 

It is worth mentioning that away from the critical point, the entanglement distance vanishes for degenerate GSs of $Z_N$ parafermion chains in the thermodynamic limit~\cite{Comment0}.

\subsection{B.  Entanglement distance in non-interacting fermion systems}

The entanglement properties of free fermions can be easily computed using the single particle correlation matrix $G_{ij}(a) \equiv {}_a\bra{\Psi}c_i^\dag c_j\ket{\Psi}_a $~\cite{Chung2001a,Cheong2004a}.
The RDM has a simple form namely: $\rho^{(a)}_L = \frac{1}{Z}\exp\para{-\sum_{i,j}h^{(a)}_{i,j}c_{i}^\dag c_{j}}$ where $\hat{h}^{\para{a}} = \log\para{\para{G^{(a)}_{LL}}^{-1} - \mathbb{1}}$, and $G^{(a)}_{LL}$ is the reduced correlation function (its submatrix). It can be shown that  
$\mathrm{tr ~}\rho_L^{\para{a}}\rho_L^{\para{b}} = \det ~\para{G_{LL}^{\para{a}} G_{LL}^{\para{b}} + \para{\mathbb{1} - G_{LL}^{\para{a}}}\para{\mathbb{1} - G_{LL}^{\para{b}}}}$. Therefore, the 2nd Renyi entropy of the many-body state $\ket{\psi}_a$ is:

\beq
S^{(a)}_{2}   ~~=  -\log~\mathrm{tr ~}\para{\rho^{(a)}_L}^2 = - \sum_{l} \log \para{p_l^2 + \para{1-p_l}^2},\label{S2}
\eeq
where $p_l$'s are eigenvalues of $G^{(a)}_{LL}$. Now, we define the truncation (projection) matrix formed of $\chi_{a}$ eigenvectors of $G_{LL}^{(a)}$ with largest $s_l \equiv -\log \para{p_l^2 + \para{1-p_l}^2}$ values. Therefore, the projected reduced correlation matrix of state $\ket{\Psi}_b$ is: $\overline{G}^{\para{b}}_{LL} = {P^{\para{a}}}^{\dag}_{L}G_{LL}^{\para{b}}P_L^{\para{a}} $. The first entanglement distance in Eq. \eqref{EPS1} can be easily computed. However, we make an indirect measurement of the second metric in Eq. \eqref{EPS2} through $\epsilon_2^{\chi_a}\para{a,b} = 1 - \bar{S}^{(b)}/S^{b}_2$ where $\bar{S}^{(b)}$ is the 2nd Renyi entropy associated with the projected reduced correlation matrix $\overline{G}^b_{LL}$. It can be numerically verified that for many-body states the two definitions of $\epsilon_2^{\chi_a}\para{a,b}$ behave similarly.  

Figure \ref{fig1} shows that for gapless fermions in 1D, the entanglement distance between $\ket{gs}$ and $c_{q}^\dag c_{k_F}\ket{gs}$ increases quadratically with the excitation energy up  to some energy index threshold equal to $\mathcal{N}/\ell$ and then starts oscillating around the saturation point with a wavelength again equal to $\mathcal{N}/\ell$. This behavior can be understood by noting that in free fermion systems all entanglement measures are deeply related to 
 $I_{m,n}\para{\ell}\equiv \int_0^{\ell} \psi^{*}_m(x)\psi_n(x)dx \propto \para{1- e^{i\para{p_m-p_n}\ell}}/\para{p_m - p_n}$ quantity where $\psi_m(x)$ is the energy eigenstate with momentum $p_m = 2\pi m/\mathcal{N}$. Apart from its envelope, $I_{m,n}\para{\ell}$ has oscillations of wavelength $\mathcal{N}/\ell$.
For massive fermions, Fig. \ref{fig3} shows that we obtain two branches both having a scaling behavior in $\ell/\mathcal{N}$ and $\Delta E$ though with different exponents. The two branches are distinguished by the momentum transfer. The branch with lower entanglement distance and $(\alpha_E,\alpha_x) = (2,4)$ exponents is related to momentum transfers less than $\pi/2$ and the remaining brach contains states with momentum transfer larger than $\pi/2$. Again, such a ramification is indeed related to a similar behavior in $I_{m,n}\para{\ell}$ for gapped systems.
Finally, for the case of massive Potts model, Fig. \ref{fig4} indicates that the entanglement distance saturates when the subsystem size becomes large compared to the correlation length. Accordingly, the entanglement distance provides an alternative way of measuring the correlation length.  

\subsection{C. Entanglement distance in 2+1 D topologically ordered states}
It is argued in Ref. \cite{Qi_Katsura2012a} that the RDM of the GS with topological charge $a$ can be obtained through boundary CFT - bulk wavefunction correspondence leading to $\rho_L^{\para{a}} \propto \hat{P}_{a}\exp(-\beta_{\rm eff} H_{\rm CFT})\hat{P}_{a}$ relation where $\hat{P}_a$ is the projector into topological sector with charge $a$. This immediately leads us to the following central result:
\beq
\mathrm{tr~}\rho_L^{\para{a}}\rho_L^{\para{b}} =\delta_{ab} \exp\para{-\alpha \ell + n_B\log\para{\mathcal{D}/d_a} },\label{Mab}
\eeq
where $\mathcal{D} = \sqrt{\sum_{a} d_a^2}$ is the total quantum dimension, $d_a$ is the quantum dimension of anyon with charge $a$, $\ell$ denotes the boundary length, $\alpha$ is a non-universal constant and $n_B$ denotes the number of boundaries (for cylinder and torus geometries we usually consider $n_B = 1$ and $ n_B=2$, respectively.) We have also used the fact that the second Renyi entropy of a 2+1 D topological state is $\alpha L - n_B\log\para{\mathcal{D}}/d_a$. We have numerically verified the predicted orthogonality for the Laughlin states~\cite{Laughlin1983a} up to $16$ electrons using ED and DMRG methods. Now it is quite easy to compute $\epsilon_1$:
\beq
\epsilon_1\para{a,b} = \para{d_a^{-n_B} + d_b^{-n_B}}.  \label{EPS4} 
\eeq
Hence, degenerate GSs have maximum distance from each other and using one of the GSs only for the truncation matrix causes a severe loss of information. However, in the next section we show that there is simple resolution of this issue at the cost of a linear increase of computation time. 
We like to emphasize that the entanglement distance of two degenerate states is maximal for 2+1D topological phases only when the two states are related by a deconfined anyon excitation (or a Wilson loop). This is not always the case. For example, parafermion zero modes/genons \cite{Clarke2012a,Lindner_Berg2013a,Cheng2012a,Vaezi_ftsc_2013a,Barkeshli_Jian2013a,You_Wen2012a} are bound to domain walls and the entanglement distance between degenerate states related through their actions vanishes. 

\section{Computational complexity of multi-state DMRG}

To simulate $n$ (nearly) degenerate quantum states, we first need to find the optimal truncation matrix $T_L$ with dimension $\mathcal{D}_L \times \chi$ that contains information about all $n$ states. The truncation error associated with the $a$-th reduced matrix is $\epsilon^{(a)} = 1- \mathrm{tr~}T_L^\dag \rho_L^{\para{a}}T_L$. Let us weight each truncation error by $p_a$ which may depend on the energy of state $a$ e.g., $p_a = e^{-\beta_{\rm eff}E_a}/Z$ or its entanglement or both. For topological states a reasonable choice can be $p_a = d_a^2/\mathcal{D}^2$. The total weighted cost function becomes: $\epsilon_{\rm eff} =  \sum_a p_a \epsilon^{(a)}$ which can be rewritten as : $\epsilon_{\rm eff} = 1- \mathrm{tr~}T_L^\dag \rho^{\rm eff}_LT_L$, where $\rho^{\rm eff}_L \equiv  \sum_a p_a \rho_L^{\para{a}}$. Therefore the optimal choice for $T_L$ minimizes the truncation error of $\rho^{\rm eff}_L$. Now we can find an estimate for the value of the effective bond dimension $\chi_{\rm eff}$. Intuitively we expect $\chi_{\rm eff} \sim \exp\para{S_2^{\rm eff}} = 1/\mathrm{tr~} \para{\rho_L^{\rm eff}~^2}$. Moreover, recall that the computation time of DMRG scales as $\chi_{\rm eff}^3$. The entanglement distance that we computed previously helps us to estimate this value. Since the GS of the system (with trivial charge for topological states) has the lowest EE among all nearby states, the EE associated with  $\rho^{\rm eff}_L$ is necessarily larger than that of the GS with trivial charge. For 1+1D systems, since the entanglement distance varies linearly as a function of energy separations and thus vanishes for degenerate states,  $\chi_{\rm eff} \sim \chi_{\mathbb{1}}$ as $\rho_{\rm eff}$ has almost the same EE as the trivial GS. However, for 2+1 D topological states, the situation is quite different. Using  Eq. \eqref{Mab} we can find $S_2^{\rm eff}$ as:
\beq
S_2^{\rm eff} = S_2\para{\rho_{\rm eff}} = \alpha \ell - n_B \log \mathcal{D} - \log \para{\sum_a \frac{p_a^2}{d_a^{n_B}}} \label{Seff}
\eeq
which is consistent with the result found in \cite{Dong_Fradkin2008a}. Therefore, $\chi_{\rm eff} = \frac{1}{\sum_{a} \frac{p_a^2}{d_a^{n_B}}}\chi_{\mathbb{1}}$. In the next section, based on the ideas developed above, we introduce two multi-state DMRG algorithms that are helpful for finding all of the (nearly) degenerate ground-states and give more accurate results for the low energy excitations. Furthermore, we verify the above relation by studying a 1/3 Laughlin state.

\section{Two DMRG-based algorithms for  obtaining degenerate ground-states}
In this section, we introduce two multi-state targeting DMRG algorithms that can be justified using the notion of entanglement distance. For a lightening introduction to the single-state DMRG algorithm see Appendix A.

\begin{figure}
\centerline{\includegraphics[width =0.9\linewidth] {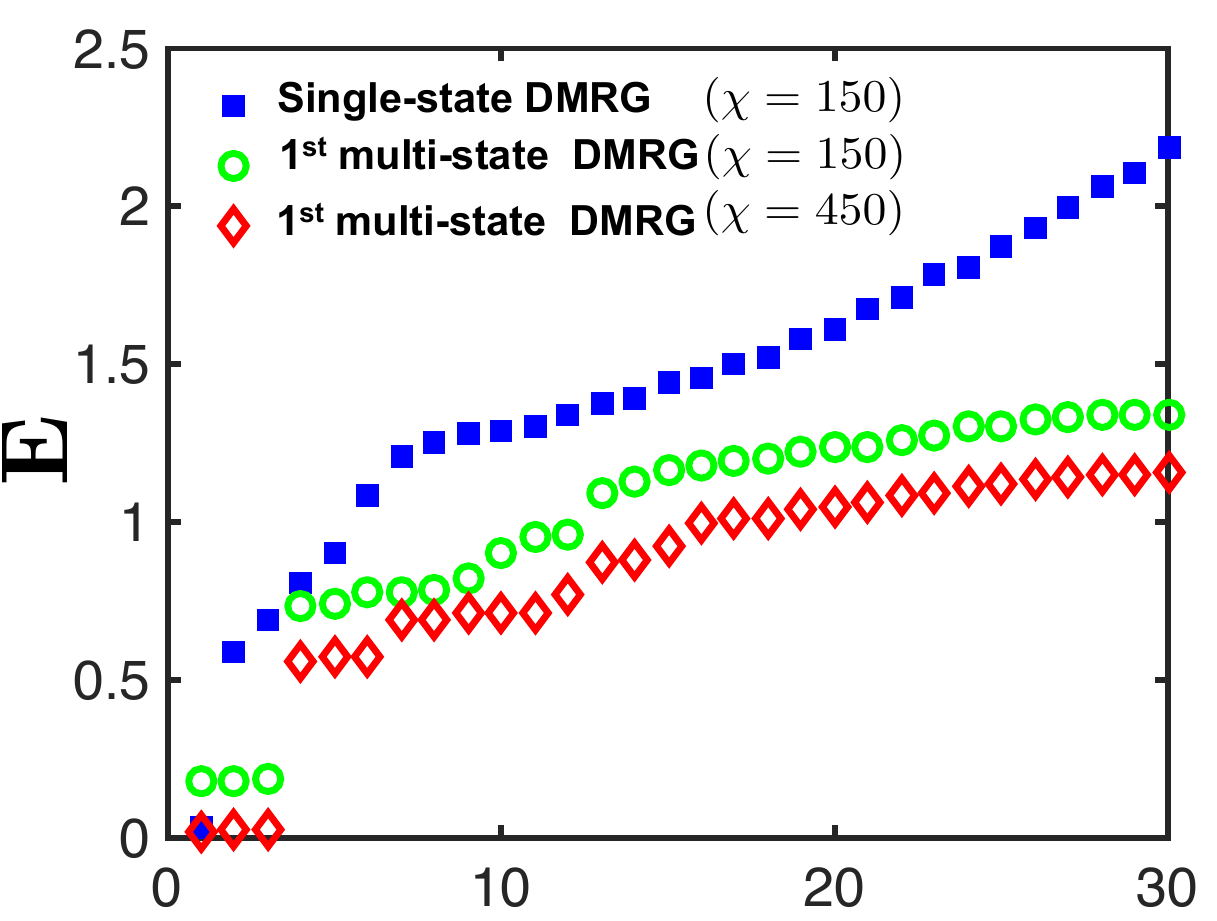}}
\caption{\label{fig5}  Comparing the results of the single-state DMRG with the first multi-state DMRG algorithm discussed in the paper for the lowest energy eigenvalues. We consider a Laughlin state at $1/3$ filling, torus geometry, with $N_e = 12$ electrons and $L_y = 15$ circumference.  We have considered Haldane's $V_1$ pseudopotential~\cite{Haldane1983a} for the interaction term. We have implemented center of mass momentum conservation mod $N_e$, thus the three topological sectors have the same total momentum (mod $N_e$). The single-state DMRG with $\chi = 150$ finds only one of the 3 degenerate ground-states (enlarging bond dimension does not help in finding more states). On the other hand, the first multi-state DMRG algorithm finds all three degenerate ground-states for both $\chi=150$, and $\chi = 450$. Although the ground-state energy of $\chi=150$ is nonzero and larger than its true value, the excitation energies ($E_i-E_0$) are estimated well. Furthermore, we see that the first multi-state DMRG with $\chi_{\rm off} = 450$ gives a ground-state energy close to zero (and that of the single-state DMRG method targeting one ground-state) consistent with our theoretical expectations (see Eq. \eqref{Chieff}).}
\end{figure}

\begin{figure}
\centerline{\includegraphics[width =0.9\linewidth] {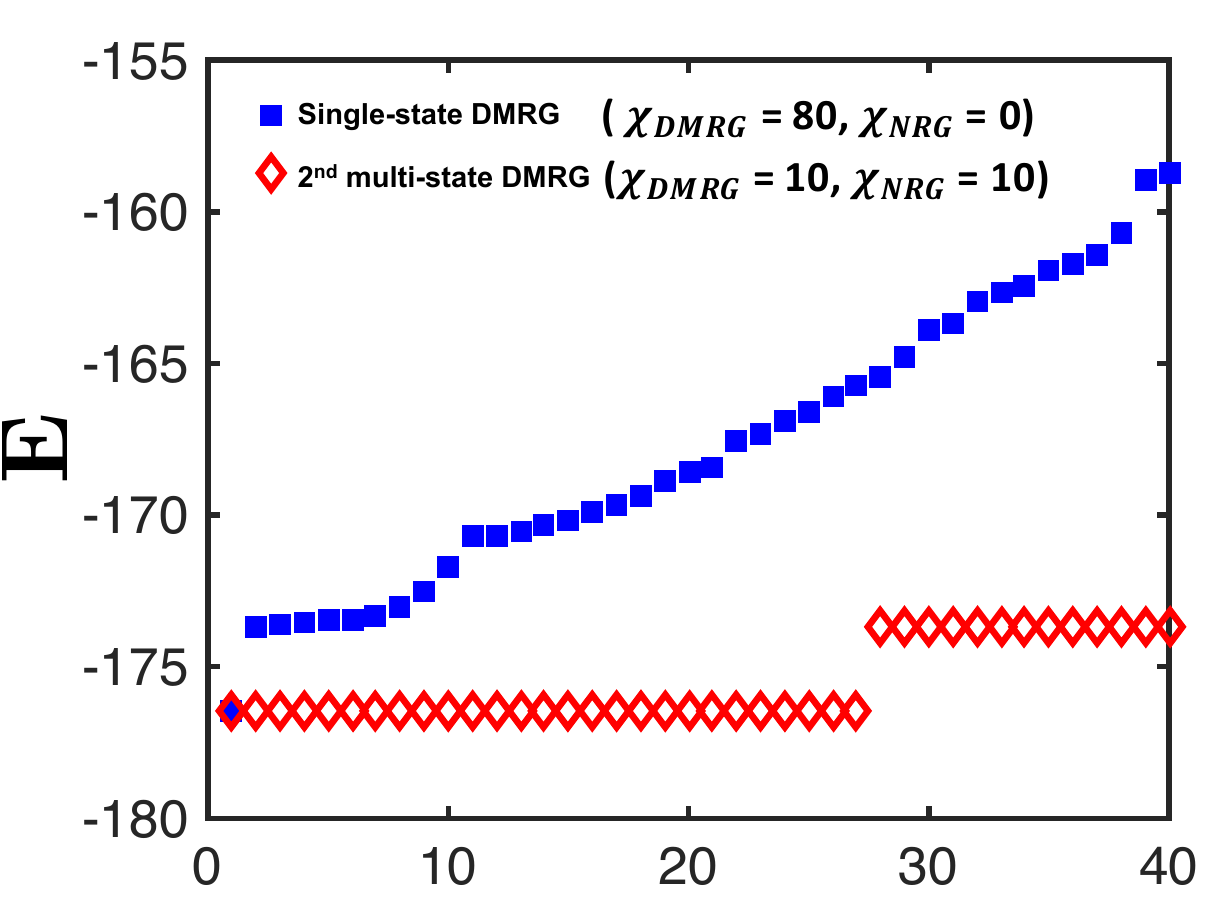}}
\caption{\label{fig6}  Comparing the results of the single-state DMRG with the second multi-state DMRG algorithm introduced in this paper for the lowest energy eigenvalues.  A $Z_3$ parafermion chain of length $\mathcal{N} = 90$ with 6 domain walls leading to $(GSD) = 27$. Again single-state DMRG cannot find all ground states even with $\chi_{\rm tot} = 80$ while the second multi-state DMRG can easily find all 27 ones with $\chi_{\rm tot} = 20$.}
\end{figure}

\subsection{Algorithm I}
Let us consider $n$ (nearly) degenerate ground-states . We want to find the optimal truncation matrix $T_L$ with dimension $\mathcal{D}_L \times \chi$ that contains information about all $n$ states. The truncation error associated with the $a$-th reduced matrix is $\epsilon^{(a)} = 1- \mathrm{tr~}T_L^\dag \rho_L^{\para{a}}T_L$. Let us weight each truncation error by $p_a$ which may depend on the energy of state $a$ e.g., $p_a = e^{-\beta_{\rm eff}E_a}/Z$ or its entanglement or both. For topological states a reasonable choice can be $p_a = d_a^2/\mathcal{D}^2$. The total weighted cost function becomes: $\epsilon_{\rm eff} =  \sum_a p_a \epsilon^{(a)}$ which can be rewritten as : $\epsilon_{\rm eff} = 1- \mathrm{tr~}T_L^\dag \rho^{\rm eff}_LT_L$, where $\rho^{\rm eff}_L \equiv  \sum_a p_a \rho_L^{\para{a}}$. Therefore the optimal choice for $T_L$ minimizes the truncation error of $\rho^{\rm eff}_L$. Thus, at each step of DMRG or related methods, we need to find $n$ lowest energy eigenstates and combine them properly to build the effective RDM and use it to achieve $T_L$ for truncation. Fulfilling this requirement, we are guaranteed to find all ground states in later steps. Otherwise, even for fairly large values of bond dimension $\chi_{\rm eff}$, the information about $n-1$ states will leak out in a rate depending on the entanglement distance and there is no guarantee that the single-state DMRG can recover all degenerate states in the later steps. Now, we need to find an estimate for the value of the effective bond dimension $\chi_{\rm eff}$. Intuitively we expect $\chi_{\rm eff} \sim \exp\para{S_2^{\rm eff}} = 1/\mathrm{tr~} \para{\rho_L^{\rm eff}~^2}$. The entanglement distance that we computed previously helps us to estimate this value. Since the ground state of the system (with trivial charge for topological states) has the lowest EE among all nearby states, the EE associated with  $\rho^{\rm eff}_L$ is necessarily larger than that of the ground state with trivial charge. For 1+1D systems, since the entanglement distance varies linearly as a function of energy separations and thus vanishes for degenerate states,  $\chi_{\rm eff} \sim \chi_{\mathbb{1}}$ as $\rho_{\rm eff}$ has almost the same EE as the trivial ground-state. However, for 2+1 D topological states, the situation is quite different. As we demonstrated in the main text, 
\beq
\mathrm{tr~}\rho_L^{\para{a}}\rho_L^{\para{b}} =\delta_{ab} \exp\para{-\alpha \ell + n_B\log\para{\mathcal{D}/d_a} },\label{Mab}
\eeq
where $\mathcal{D} = \sqrt{\sum_{a} d_a^2}$ is the total quantum dimension, $d_a$ is the quantum dimension of anyon with charge $a$, $\ell$ denotes the boundary length, $\alpha$ is a non-universal constant and $n_B$ denotes the number of boundaries (for cylinder and torus geometries we usually consider $n_B = 1$ and $ n_B=2$, respectively.) Using  Eq. \eqref{Mab} we can find $S_2^{\rm eff}$ as:
\beq
S_2^{\rm eff} = S_2\para{\rho_{\rm eff}} = \alpha \ell - n_B \log \mathcal{D} - \log \para{\sum_a \frac{p_a^2}{d_a^{n_B}}} \label{Seff}.
\eeq

Therefore, 
\beq
\chi_{\rm eff} = \frac{1}{\sum_{a} \frac{p_a^2}{d_a^{n_B}}}\chi_{\mathbb{1}}.  \label{Chieff}
\eeq
Note that in this expression for $\chi_{\rm eff}$ we must include {\em deconfined anyons} only. Fig. \ref{fig5} shows how this modified DMRG achieves the correct GSD for Laughlin states. We would like to mention that although similar methods have been used in the past to find the low energy excitations as well as ground-state degeneracy, we add an important flavor to it, namely we increase the system by a specific number related to the topological order of that phase at each step of the infinite DMRG. Without this seemingly simple modification, there is no guarantee to find all of the degenerate states as one can verify it for simple Hamiltonian. For 2D topological states, the number of sites added must be equal to the size of the unit cell in the thin torus pattern of that phase. This is something which was missed in the previous studies, and its importance can be understood as follows.  The mentioned multi-state DMRG algorithm requires finding all degenerate ground-states at each step of DMRG e.g., via Lanczos method. This increases the computation time unless we optimize  the procedure by modifying the wave-function transformation and Lanczos  algorithm to use $n$ initial vectors or a linear combination of them. Secondly, for the case of topological states (e.g., Laughlin states at filling $1/m$ on torus geometry), the GSD is finite ($m$-fold degenerate for $1/m$ Laughlin state) only if the total system size has certain length (multiples of $m$ for $1/m$ Laughlin state), otherwise it can grow polynomially in the system size (or more precisely given by the quasi-hole counting of that state). As a result, the required bond dimension to keep the truncation error small explodes, and the quantum simulation becomes intractable after a few steps. Therefore, in order to resolve this severe limitation, we must increase the system size during the infinite DMRG steps such that the GSD remains constant. For example, we must add $m$ or multiples of $m$ sites at each step of the infinite DMRG for Laughlin states at $1/m$ filling. Therefore, the dimension of the Hilbert space and operators will keep growing until the $m$-th site is added after which we truncate the operators and states to reduce their dimensions down to $\chi_{\rm eff}$. For the finite DMRG part, we no longer have this issue because the system size if fixed. Thus,
 one site can be added and removed since the total system size is fixed. Furthermore, in order to reduce the number of iterations in diagonalizing the super-block Hamiltonian we suggest utilizing the Arnoldi method instead of the Lanczos method and use all of the degenerate wave-functions states from the previous step as the initial vectors spanning the Krylov subspace (multi-state wave function transformation).

\subsection{Algorithm II}

The second algorithm that we introduce for obtaining the nearly degenerate ground-states (as well as low-lying excitations) combines ideas from White's DMRG and Wilson's numerical RG (NRG)~\cite{Wilson1975a} methods. Notably, NRG does not explicitly break ergodicity since truncation matrices are obtained from sub-system Hamiltonian instead of RDMs. However, except for certain systems it usually provides an unsatisfactory estimate of the ground-state energy. On the other hand, the White's DMRG estimates energy very well but cannot keep track of all degenerate states. Now let us consider 
$T_L^{\rm DMRG}$ with dimension $\mathcal{D}_L \times \chi_{\rm DMRG}$ obtained from the RDM in the usual way, and $T_L^{\rm NRG}$ with dimension $\mathcal{D}_L \times \chi_{\rm NRG}$ that is obtained by putting lowest $\chi_{\rm NRG}$ eigenstates of left Hamiltonian $H_L$ together.  The truncation matrix is obtained via the concatenation of these two truncation matrices, $T_L = \left[T_L^{\rm DMRG}~~T_L^{\rm NRG}\right]$, followed by orthogonalization of the two sub-matrices to enforce $T_L^\dag T_L = \mathbb{1}$ constraint. The computation cost of this is same as that of DMRG with $\chi = \chi_{\rm DMRG} + \chi_{\rm NRG}$. This method can obtain all degenerate states of 1+1D systems. For example, consider $Z_3$ parafermion chain with $6$ domain walls and $J_{FM} = h_{PM} = 1$ and $J_{PM} = h_{PM} = 0.1e^{i\pi/10}$ whose $GSD = 27$. The ground-state energy for these parameters can be obtained using the single-state DMRG with $\chi_{\rm DMRG} \sim 5$. However, we just obtain one ground state  even if we consider $\chi_{\rm DMRG} = 80$. Nonetheless, with our second multi-state DMRG, $\chi_{\rm DMRG} = 10$, and $\chi_{\rm NRG} = 10$ are sufficient to obtain all degenerate states. (see Fig \ref{fig6}). It is worth noting that even within the single-state DMRG algorithm, one can find $N$ ground states among possibly more degenerate ground-states of $Z_N$ parafermion chains through implementing $Z_N$ symmetry. Yet, the remaining degenerate ground-states (see Fig. \ref{fig6} for example) cannot be obtained in the single-state DMRG.

{\bf  {\em Note added.--} } After the completion of this work, we became aware of a recent related work on $\epsilon_1\para{a,b}$ in the 1+1D CFTs~\cite{Ugajin2016a} with results similar to ours.

\begin{acknowledgements}
\section*{Acknowledgements} 
We gratefully acknowledge helpful discussions with Zohar Nussinov, Hongchen Jiang, Shenxiu Liu, Xiao-Liang Qi, Pavan Hosur, Gerardo Ortiz and Xueda Wen. A. V. was funded by the Gordon and Betty Moore Foundation?s EPiQS Initiative through Grant GBMF4302 and M.-S. V. was supported by the National Science Foundation under NSF Grant No. DMR-1411229. \\

\end{acknowledgements}


\section*{Appendix A :  A brief review of the single-state DMRG algorithm}

The basic idea behind the DMRG approach is the singular-value decomposition (SVD) of the ground-state. Consider a quantum system partitioned into $L$ and $R$ subsystems with Hilbert spaces of dimensions $\mathcal{D}_L$ and $\mathcal{D}_R$, respectively. The ground-state wave-function is a $\mathcal{D}_L\mathcal{D}_R$ dimensional vector which can be reshaped into a $\mathcal{D}_L \times \mathcal{D}_R$ dimensional matrix, $\Psi_{ij}$, where $i = 1\cdots \mathcal{D}_L$ and $j = 1\cdots \mathcal{D}_R$. The single value decomposing states that $\Psi$ can be rewritten as $\Psi = U\Lambda^{1/2} V^{\rm T}$, where $U$ ~($V$) is a unitary matrix formed of juxtaposing the eigenstates of $\rho_L = \Psi \Psi^\dag$ ~($\rho_R = \Psi^{\rm T} \Psi^*$), and $\Lambda$ is a diagonal matrix formed of eigenvalues of $\rho_L$ or $\rho_R$. It is easy to show that $\rho_L$ and $\rho_R$ are indeed the RDMs associated with the left and right subsystems respectively. Similarly, the EE associated with the chosen partitioning is $S = - \mathrm{tr~} \Lambda \log \Lambda$. For local gapped Hamiltonians, the SVD (a.k.a. Schmidt decomposition) of the ground-state is much less complex than a generic excited state, namely most of the diagonal elements of the eigenvalue matrix $\Lambda$ are negligible leading to a lower EE for the ground-state. This allows us to use an efficient principal component analysis by keeping eigenvalues larger than a threshold, $\lambda_{th}$. Let us assume the number of eigenvalues satisfying this condition is $\chi$. Accordingly, instead of $U$ ~($V$) we must use $T_L$~($T_R$) which contains the $\chi$ dominant eigenvectors of $\rho_L$ ~$(\rho_R)$. We also have $T_L^\dag T_L = \mathbb{1}$ and a similar relation for $T_R$. So, $T_L$~ ($T_R$) is a projection (truncation) operators that can project ground-state and operators defined in the  $L$~($R$) subsystem to the subspace spanned by important eigenstates of the density matrices. For example, we expect $\overline{\Psi} = T_L^\dag \Psi T_R^{*}$ to contain almost all of the information stored in $\Psi$ e.g., we can use it to find the correlation function, entanglement, ground-state energy and etc. Therefore, the error of calculating these quantities w.r.t. $\overline{\Psi}$ decays exponentially by increasing $\chi$.

\begin{figure*}
\centerline{\includegraphics[width  =0.9\linewidth] {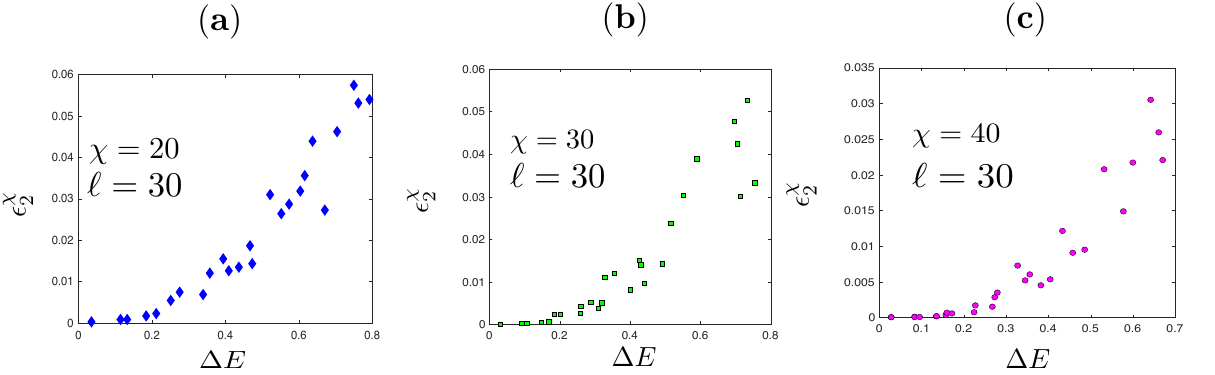}}
\caption{\label{figA3}  DMRG results for $\epsilon_2^{\chi}$ in a critical $Z_3$ parafermion chain of $N=100$ total length measured at $\ell = 30$ for three different values of bond dimension, $\chi$.  These results suggest that $\beta_E(\chi) = \chi/20 + 1.3$  ($\ell = 30$) with a high accuracy.}
\end{figure*}

In practice, since we do not know the wavefunction a priori and thus we cannot find the truncation matrices, we need to find an efficient way of achieving them. DMRG provides one way of reaching this goal, though in most 2+1D systems the bond dimension, $\chi$, has an exponential dependence on the system width (due to the area law EE in real space), hence we can only simulate narrow systems. The DMRG algorithm finds $T_L$ and $T_R$ by iteration. Instead of considering the whole system of $\mathcal{N}$ sites (where the dimension of local (onsite) Hilbert space is $d$), in the $n-th$ step of iteration, DMRG partitions the system into three subregions, left, middle and right  with $n$, $\mathcal{N}-2n$ and $n$ sites respectively. Then it assumes region $M$ is decoupled from the rest and also $L$ and $R$ regions interact with each other directly as if they are neighbors and attached (obviously these assumptions generate some errors and DMRG needs to fix them in later steps).  For small $n$ we can use ED to find the ground-state(s) exactly and no truncation is needed. The Hamiltonian of the $L$ and $R$ regions can in general be written as:
\beq
H_{LR} = H_L(n) \otimes \mathbb{1}_R + \mathbb{1}_L\otimes H_R (n)+ \sum_i g_i O_L^i(n)\otimes O_R^i(n),
\eeq
where $g_i$ are coupling constants. The soon as the Hilbert space dimensions of the $L$ or $R$ regions exceed $\chi$, we start truncating operators by $T_L(n)$ and $T_R(n)$ obtained as described above, after which we obtain:

\beq
\overline{H_{LR}} = \overline{H_L}(n) \otimes \overline{\mathbb{1}_R} + \overline{\mathbb{1}_L}\otimes \overline{H_R}(n) + \sum_i g_i \overline{O_L}^i(n)\otimes \overline{O_R}^i(n),
\eeq
where $\overline{O_L}^i(n) = T_L^\dag(n) O_L^i(n) T_L(n)$, and a similar expression for $\overline{O_R}^i(n)$. The truncated operators are $\chi \times \chi$ dimensional. In step $n+1$, DMRG adds one site to the left and one site to the right, after which the dimension of the left (and also right) region is $d\chi$. Again, we construct the total Hamiltonian, find its ground-state and obtain the truncation matrices for step $n+1$, $T_L(n+1)$ and $T_R(n+1)$. Then we use the truncation matrices to truncate the operators once more after which their dimensions reduce to $\chi \times \chi$ again. This way, the dimension of operators is kept constant instead of growing exponentially with $n$. This procedure of adding sites followed by truncations is repeated until $n = \mathcal{N}/2$. So far, we assumed that only $n+n$ sites in the $L$ and $R$ regions interact, and therefore $\ket{\Psi}_{L,n;M,\mathcal{N}-2n;R,n} = \ket{\Psi }_{L,n;R,n}\otimes \ket{\Psi}_{M,\mathcal{N} - 2n}$ which except for gapped $1+1D$ systems is not a good approximation. In order to improve this assumption, DMRG uses the so-called sweeps, where region $M$ disappears and in step $n$, the left region contains $n$ sites and the right region contains the remaining $\mathcal{N} -n$ sites. The truncation matrices and operator representations for $(L,n)$ and $(R,\mathcal{N}-n)$ from previous iterations are used. Then we construct the total Hamiltonian, find the ground-state and use them to update $T_L(n)$, and $T_R(\mathcal{N}-n)$ as well as operators $\overline{O_L}^i(n)$ and $\overline{O_R}^i(\mathcal{N} - n)$. After a few sweeps across the system, the algorithm converges and the ground-state energy, correlation functions and etc. can be obtained with a high accuracy for large enough values of $\chi$. One criterion is that $\chi$ must be the order of ${\rm max}_n(e^{S(n)})$ at least, where $S(n)$ is the EE associated with $n$ sites in the left.

\section*{Appendix B : More results for critical $Z_3$ clock model}
In Fig. \ref{figA3}, we present more results for $\epsilon_2^{\chi}$ measured at $\ell = 30$ for a critical $Z_3$ parafermion chain of $N=100$ sites. Using a polynomial fit,  one can easily verify the expression in the caption of Fig. 2 of the main text: $\beta_E(\chi) = \chi/20 + 1.3$  ($\ell = 30$).


\begin{thebibliography}{57}%
\makeatletter
\providecommand \@ifxundefined [1]{%
 \@ifx{#1\undefined}
}%
\providecommand \@ifnum [1]{%
 \ifnum #1\expandafter \@firstoftwo
 \else \expandafter \@secondoftwo
 \fi
}%
\providecommand \@ifx [1]{%
 \ifx #1\expandafter \@firstoftwo
 \else \expandafter \@secondoftwo
 \fi
}%
\providecommand \natexlab [1]{#1}%
\providecommand \enquote  [1]{``#1''}%
\providecommand \bibnamefont  [1]{#1}%
\providecommand \bibfnamefont [1]{#1}%
\providecommand \citenamefont [1]{#1}%
\providecommand \href@noop [0]{\@secondoftwo}%
\providecommand \href [0]{\begingroup \@sanitize@url \@href}%
\providecommand \@href[1]{\@@startlink{#1}\@@href}%
\providecommand \@@href[1]{\endgroup#1\@@endlink}%
\providecommand \@sanitize@url [0]{\catcode `\\12\catcode `\$12\catcode
  `\&12\catcode `\#12\catcode `\^12\catcode `\_12\catcode `\%12\relax}%
\providecommand \@@startlink[1]{}%
\providecommand \@@endlink[0]{}%
\providecommand \url  [0]{\begingroup\@sanitize@url \@url }%
\providecommand \@url [1]{\endgroup\@href {#1}{\urlprefix }}%
\providecommand \urlprefix  [0]{URL }%
\providecommand \Eprint [0]{\href }%
\providecommand \doibase [0]{http://dx.doi.org/}%
\providecommand \selectlanguage [0]{\@gobble}%
\providecommand \bibinfo  [0]{\@secondoftwo}%
\providecommand \bibfield  [0]{\@secondoftwo}%
\providecommand \translation [1]{[#1]}%
\providecommand \BibitemOpen [0]{}%
\providecommand \bibitemStop [0]{}%
\providecommand \bibitemNoStop [0]{.\EOS\space}%
\providecommand \EOS [0]{\spacefactor3000\relax}%
\providecommand \BibitemShut  [1]{\csname bibitem#1\endcsname}%
\let\auto@bib@innerbib\@empty
\bibitem [{\citenamefont {White}(1992)}]{White1992a}%
  \BibitemOpen
  \bibfield  {author} {\bibinfo {author} {\bibfnamefont {S.~R.}\ \bibnamefont
  {White}},\ }\href {\doibase 10.1103/PhysRevLett.69.2863} {\bibfield
  {journal} {\bibinfo  {journal} {Phys. Rev. Lett.}\ }\textbf {\bibinfo
  {volume} {69}},\ \bibinfo {pages} {2863} (\bibinfo {year}
  {1992})}\BibitemShut {NoStop}%
\bibitem [{\citenamefont {\"Ostlund}\ and\ \citenamefont
  {Rommer}(1995)}]{Ostlund1995a}%
  \BibitemOpen
  \bibfield  {author} {\bibinfo {author} {\bibfnamefont {S.}~\bibnamefont
  {\"Ostlund}}\ and\ \bibinfo {author} {\bibfnamefont {S.}~\bibnamefont
  {Rommer}},\ }\href {\doibase 10.1103/PhysRevLett.75.3537} {\bibfield
  {journal} {\bibinfo  {journal} {Phys. Rev. Lett.}\ }\textbf {\bibinfo
  {volume} {75}},\ \bibinfo {pages} {3537} (\bibinfo {year}
  {1995})}\BibitemShut {NoStop}%
\bibitem [{\citenamefont {Vidal}(2007)}]{Vidal2007a}%
  \BibitemOpen
  \bibfield  {author} {\bibinfo {author} {\bibfnamefont {G.}~\bibnamefont
  {Vidal}},\ }\href {\doibase 10.1103/PhysRevLett.99.220405} {\bibfield
  {journal} {\bibinfo  {journal} {Phys. Rev. Lett.}\ }\textbf {\bibinfo
  {volume} {99}},\ \bibinfo {pages} {220405} (\bibinfo {year}
  {2007})}\BibitemShut {NoStop}%
\bibitem [{\citenamefont {Verstraete}\ \emph {et~al.}(2008)\citenamefont
  {Verstraete}, \citenamefont {Murg},\ and\ \citenamefont
  {Cirac}}]{Cirac2008a}%
  \BibitemOpen
  \bibfield  {author} {\bibinfo {author} {\bibfnamefont {F.}~\bibnamefont
  {Verstraete}}, \bibinfo {author} {\bibfnamefont {V.}~\bibnamefont {Murg}}, \
  and\ \bibinfo {author} {\bibfnamefont {J.}~\bibnamefont {Cirac}},\ }\href
  {\doibase 10.1080/14789940801912366} {\bibfield  {journal} {\bibinfo
  {journal} {Advances in Physics}\ }\textbf {\bibinfo {volume} {57}},\ \bibinfo
  {pages} {143} (\bibinfo {year} {2008})},\ \Eprint
  {http://arxiv.org/abs/http://dx.doi.org/10.1080/14789940801912366}
  {http://dx.doi.org/10.1080/14789940801912366} \BibitemShut {NoStop}%
\bibitem [{\citenamefont {{Schollw{\"o}ck}}(2011)}]{Schollwock2011a}%
  \BibitemOpen
  \bibfield  {author} {\bibinfo {author} {\bibfnamefont {U.}~\bibnamefont
  {{Schollw{\"o}ck}}},\ }\href {\doibase 10.1016/j.aop.2010.09.012} {\bibfield
  {journal} {\bibinfo  {journal} {Annals of Physics}\ }\textbf {\bibinfo
  {volume} {326}},\ \bibinfo {pages} {96} (\bibinfo {year} {2011})},\ \Eprint
  {http://arxiv.org/abs/1008.3477} {arXiv:1008.3477 [cond-mat.str-el]}
  \BibitemShut {NoStop}%
\bibitem [{\citenamefont {Stoudenmire}\ and\ \citenamefont
  {White}(2012)}]{Stoudenmire_2012a}%
  \BibitemOpen
  \bibfield  {author} {\bibinfo {author} {\bibfnamefont {E.}~\bibnamefont
  {Stoudenmire}}\ and\ \bibinfo {author} {\bibfnamefont {S.~R.}\ \bibnamefont
  {White}},\ }\href {\doibase 10.1146/annurev-conmatphys-020911-125018}
  {\bibfield  {journal} {\bibinfo  {journal} {Annual Review of Condensed Matter
  Physics}\ }\textbf {\bibinfo {volume} {3}},\ \bibinfo {pages} {111} (\bibinfo
  {year} {2012})},\ \Eprint
  {http://arxiv.org/abs/http://dx.doi.org/10.1146/annurev-conmatphys-020911-125018}
  {http://dx.doi.org/10.1146/annurev-conmatphys-020911-125018} \BibitemShut
  {NoStop}%
\bibitem [{\citenamefont {Vidal}(2004)}]{Vidal2004a}%
  \BibitemOpen
  \bibfield  {author} {\bibinfo {author} {\bibfnamefont {G.}~\bibnamefont
  {Vidal}},\ }\href {\doibase 10.1103/PhysRevLett.93.040502} {\bibfield
  {journal} {\bibinfo  {journal} {Phys. Rev. Lett.}\ }\textbf {\bibinfo
  {volume} {93}},\ \bibinfo {pages} {040502} (\bibinfo {year}
  {2004})}\BibitemShut {NoStop}%
\bibitem [{\citenamefont {White}\ and\ \citenamefont
  {Feiguin}(2004)}]{White_Feiguin2004a}%
  \BibitemOpen
  \bibfield  {author} {\bibinfo {author} {\bibfnamefont {S.~R.}\ \bibnamefont
  {White}}\ and\ \bibinfo {author} {\bibfnamefont {A.~E.}\ \bibnamefont
  {Feiguin}},\ }\href {\doibase 10.1103/PhysRevLett.93.076401} {\bibfield
  {journal} {\bibinfo  {journal} {Phys. Rev. Lett.}\ }\textbf {\bibinfo
  {volume} {93}},\ \bibinfo {pages} {076401} (\bibinfo {year}
  {2004})}\BibitemShut {NoStop}%
\bibitem [{\citenamefont {Daley}\ \emph {et~al.}(2004)\citenamefont {Daley},
  \citenamefont {Kollath}, \citenamefont {Schollwöck},\ and\ \citenamefont
  {Vidal}}]{Vidal2004b}%
  \BibitemOpen
  \bibfield  {author} {\bibinfo {author} {\bibfnamefont {A.~J.}\ \bibnamefont
  {Daley}}, \bibinfo {author} {\bibfnamefont {C.}~\bibnamefont {Kollath}},
  \bibinfo {author} {\bibfnamefont {U.}~\bibnamefont {Schollwöck}}, \ and\
  \bibinfo {author} {\bibfnamefont {G.}~\bibnamefont {Vidal}},\ }\href
  {http://stacks.iop.org/1742-5468/2004/i=04/a=P04005} {\bibfield  {journal}
  {\bibinfo  {journal} {Journal of Statistical Mechanics: Theory and
  Experiment}\ }\textbf {\bibinfo {volume} {2004}},\ \bibinfo {pages} {P04005}
  (\bibinfo {year} {2004})}\BibitemShut {NoStop}%
\bibitem [{\citenamefont {White}(1996)}]{White1996a}%
  \BibitemOpen
  \bibfield  {author} {\bibinfo {author} {\bibfnamefont {S.~R.}\ \bibnamefont
  {White}},\ }\href {\doibase 10.1103/PhysRevLett.77.3633} {\bibfield
  {journal} {\bibinfo  {journal} {Phys. Rev. Lett.}\ }\textbf {\bibinfo
  {volume} {77}},\ \bibinfo {pages} {3633} (\bibinfo {year}
  {1996})}\BibitemShut {NoStop}%
\bibitem [{\citenamefont {{McCulloch}}(2008)}]{McCulloch2008a}%
  \BibitemOpen
  \bibfield  {author} {\bibinfo {author} {\bibfnamefont {I.~P.}\ \bibnamefont
  {{McCulloch}}},\ }\href@noop {} {\bibfield  {journal} {\bibinfo  {journal}
  {ArXiv e-prints}\ } (\bibinfo {year} {2008})},\ \Eprint
  {http://arxiv.org/abs/0804.2509} {arXiv:0804.2509 [cond-mat.str-el]}
  \BibitemShut {NoStop}%
\bibitem [{\citenamefont {White}(2005)}]{White2005a}%
  \BibitemOpen
  \bibfield  {author} {\bibinfo {author} {\bibfnamefont {S.~R.}\ \bibnamefont
  {White}},\ }\href {\doibase 10.1103/PhysRevB.72.180403} {\bibfield  {journal}
  {\bibinfo  {journal} {Phys. Rev. B}\ }\textbf {\bibinfo {volume} {72}},\
  \bibinfo {pages} {180403} (\bibinfo {year} {2005})}\BibitemShut {NoStop}%
\bibitem [{\citenamefont {Eisert}\ \emph {et~al.}(2010)\citenamefont {Eisert},
  \citenamefont {Cramer},\ and\ \citenamefont {Plenio}}]{Eisert2010a}%
  \BibitemOpen
  \bibfield  {author} {\bibinfo {author} {\bibfnamefont {J.}~\bibnamefont
  {Eisert}}, \bibinfo {author} {\bibfnamefont {M.}~\bibnamefont {Cramer}}, \
  and\ \bibinfo {author} {\bibfnamefont {M.~B.}\ \bibnamefont {Plenio}},\
  }\href {\doibase 10.1103/RevModPhys.82.277} {\bibfield  {journal} {\bibinfo
  {journal} {Rev. Mod. Phys.}\ }\textbf {\bibinfo {volume} {82}},\ \bibinfo
  {pages} {277} (\bibinfo {year} {2010})}\BibitemShut {NoStop}%
\bibitem [{\citenamefont {{Holzhey}}\ \emph {et~al.}(1994)\citenamefont
  {{Holzhey}}, \citenamefont {{Larsen}},\ and\ \citenamefont
  {{Wilczek}}}]{Holzhey1994a}%
  \BibitemOpen
  \bibfield  {author} {\bibinfo {author} {\bibfnamefont {C.}~\bibnamefont
  {{Holzhey}}}, \bibinfo {author} {\bibfnamefont {F.}~\bibnamefont {{Larsen}}},
  \ and\ \bibinfo {author} {\bibfnamefont {F.}~\bibnamefont {{Wilczek}}},\
  }\href {\doibase 10.1016/0550-3213(94)90402-2} {\bibfield  {journal}
  {\bibinfo  {journal} {Nuclear Physics B}\ }\textbf {\bibinfo {volume}
  {424}},\ \bibinfo {pages} {443} (\bibinfo {year} {1994})},\ \Eprint
  {http://arxiv.org/abs/hep-th/9403108} {hep-th/9403108} \BibitemShut {NoStop}%
\bibitem [{\citenamefont {{Calabrese}}\ and\ \citenamefont
  {{Cardy}}(2004)}]{Calabrese2004a}%
  \BibitemOpen
  \bibfield  {author} {\bibinfo {author} {\bibfnamefont {P.}~\bibnamefont
  {{Calabrese}}}\ and\ \bibinfo {author} {\bibfnamefont {J.}~\bibnamefont
  {{Cardy}}},\ }\href {\doibase 10.1088/1742-5468/2004/06/P06002} {\bibfield
  {journal} {\bibinfo  {journal} {Journal of Statistical Mechanics: Theory and
  Experiment}\ }\textbf {\bibinfo {volume} {6}},\ \bibinfo {pages} {06002}
  (\bibinfo {year} {2004})},\ \Eprint {http://arxiv.org/abs/hep-th/0405152}
  {hep-th/0405152} \BibitemShut {NoStop}%
\bibitem [{\citenamefont {Kitaev}\ and\ \citenamefont
  {Preskill}(2006)}]{Kitaev2004a}%
  \BibitemOpen
  \bibfield  {author} {\bibinfo {author} {\bibfnamefont {A.}~\bibnamefont
  {Kitaev}}\ and\ \bibinfo {author} {\bibfnamefont {J.}~\bibnamefont
  {Preskill}},\ }\href {\doibase 10.1103/PhysRevLett.96.110404} {\bibfield
  {journal} {\bibinfo  {journal} {Phys. Rev. Lett.}\ }\textbf {\bibinfo
  {volume} {96}},\ \bibinfo {pages} {110404} (\bibinfo {year}
  {2006})}\BibitemShut {NoStop}%
\bibitem [{\citenamefont {Levin}\ and\ \citenamefont {Wen}(2006)}]{Levin2004a}%
  \BibitemOpen
  \bibfield  {author} {\bibinfo {author} {\bibfnamefont {M.}~\bibnamefont
  {Levin}}\ and\ \bibinfo {author} {\bibfnamefont {X.-G.}\ \bibnamefont
  {Wen}},\ }\href {\doibase 10.1103/PhysRevLett.96.110405} {\bibfield
  {journal} {\bibinfo  {journal} {Phys. Rev. Lett.}\ }\textbf {\bibinfo
  {volume} {96}},\ \bibinfo {pages} {110405} (\bibinfo {year}
  {2006})}\BibitemShut {NoStop}%
\bibitem [{\citenamefont {Li}\ and\ \citenamefont
  {Haldane}(2008)}]{Li_Haldane2008a}%
  \BibitemOpen
  \bibfield  {author} {\bibinfo {author} {\bibfnamefont {H.}~\bibnamefont
  {Li}}\ and\ \bibinfo {author} {\bibfnamefont {F.~D.~M.}\ \bibnamefont
  {Haldane}},\ }\href {\doibase 10.1103/PhysRevLett.101.010504} {\bibfield
  {journal} {\bibinfo  {journal} {Phys. Rev. Lett.}\ }\textbf {\bibinfo
  {volume} {101}},\ \bibinfo {pages} {010504} (\bibinfo {year}
  {2008})}\BibitemShut {NoStop}%
\bibitem [{\citenamefont {Pollmann}\ \emph {et~al.}(2010)\citenamefont
  {Pollmann}, \citenamefont {Turner}, \citenamefont {Berg},\ and\ \citenamefont
  {Oshikawa}}]{Pollmann2010a}%
  \BibitemOpen
  \bibfield  {author} {\bibinfo {author} {\bibfnamefont {F.}~\bibnamefont
  {Pollmann}}, \bibinfo {author} {\bibfnamefont {A.~M.}\ \bibnamefont
  {Turner}}, \bibinfo {author} {\bibfnamefont {E.}~\bibnamefont {Berg}}, \ and\
  \bibinfo {author} {\bibfnamefont {M.}~\bibnamefont {Oshikawa}},\ }\href
  {\doibase 10.1103/PhysRevB.81.064439} {\bibfield  {journal} {\bibinfo
  {journal} {Phys. Rev. B}\ }\textbf {\bibinfo {volume} {81}},\ \bibinfo
  {pages} {064439} (\bibinfo {year} {2010})}\BibitemShut {NoStop}%
\bibitem [{\citenamefont {Sterdyniak}\ \emph {et~al.}(2011)\citenamefont
  {Sterdyniak}, \citenamefont {Regnault},\ and\ \citenamefont
  {Bernevig}}]{Sterdyniak2011a}%
  \BibitemOpen
  \bibfield  {author} {\bibinfo {author} {\bibfnamefont {A.}~\bibnamefont
  {Sterdyniak}}, \bibinfo {author} {\bibfnamefont {N.}~\bibnamefont
  {Regnault}}, \ and\ \bibinfo {author} {\bibfnamefont {B.~A.}\ \bibnamefont
  {Bernevig}},\ }\href {\doibase 10.1103/PhysRevLett.106.100405} {\bibfield
  {journal} {\bibinfo  {journal} {Phys. Rev. Lett.}\ }\textbf {\bibinfo
  {volume} {106}},\ \bibinfo {pages} {100405} (\bibinfo {year}
  {2011})}\BibitemShut {NoStop}%
\bibitem [{\citenamefont {Zhang}\ \emph {et~al.}(2012)\citenamefont {Zhang},
  \citenamefont {Grover}, \citenamefont {Turner}, \citenamefont {Oshikawa},\
  and\ \citenamefont {Vishwanath}}]{Zhang_Grover2012a}%
  \BibitemOpen
  \bibfield  {author} {\bibinfo {author} {\bibfnamefont {Y.}~\bibnamefont
  {Zhang}}, \bibinfo {author} {\bibfnamefont {T.}~\bibnamefont {Grover}},
  \bibinfo {author} {\bibfnamefont {A.}~\bibnamefont {Turner}}, \bibinfo
  {author} {\bibfnamefont {M.}~\bibnamefont {Oshikawa}}, \ and\ \bibinfo
  {author} {\bibfnamefont {A.}~\bibnamefont {Vishwanath}},\ }\href {\doibase
  10.1103/PhysRevB.85.235151} {\bibfield  {journal} {\bibinfo  {journal} {Phys.
  Rev. B}\ }\textbf {\bibinfo {volume} {85}},\ \bibinfo {pages} {235151}
  (\bibinfo {year} {2012})}\BibitemShut {NoStop}%
\bibitem [{\citenamefont {Jiang}\ \emph {et~al.}(2008)\citenamefont {Jiang},
  \citenamefont {Weng},\ and\ \citenamefont {Sheng}}]{Jiang_2008a}%
  \BibitemOpen
  \bibfield  {author} {\bibinfo {author} {\bibfnamefont {H.~C.}\ \bibnamefont
  {Jiang}}, \bibinfo {author} {\bibfnamefont {Z.~Y.}\ \bibnamefont {Weng}}, \
  and\ \bibinfo {author} {\bibfnamefont {D.~N.}\ \bibnamefont {Sheng}},\ }\href
  {\doibase 10.1103/PhysRevLett.101.117203} {\bibfield  {journal} {\bibinfo
  {journal} {Phys. Rev. Lett.}\ }\textbf {\bibinfo {volume} {101}},\ \bibinfo
  {pages} {117203} (\bibinfo {year} {2008})}\BibitemShut {NoStop}%
\bibitem [{\citenamefont {Jiang}\ \emph {et~al.}(2012)\citenamefont {Jiang},
  \citenamefont {Wang},\ and\ \citenamefont {Balents}}]{Jiang2012a}%
  \BibitemOpen
  \bibfield  {author} {\bibinfo {author} {\bibfnamefont {H.-C.}\ \bibnamefont
  {Jiang}}, \bibinfo {author} {\bibfnamefont {Z.}~\bibnamefont {Wang}}, \ and\
  \bibinfo {author} {\bibfnamefont {L.}~\bibnamefont {Balents}},\ }\href@noop
  {} {\bibfield  {journal} {\bibinfo  {journal} {Nature Phys.}\ }\textbf
  {\bibinfo {volume} {8}},\ \bibinfo {pages} {902} (\bibinfo {year}
  {2012})}\BibitemShut {NoStop}%
\bibitem [{\citenamefont {Depenbrock}\ \emph {et~al.}(2012)\citenamefont
  {Depenbrock}, \citenamefont {McCulloch},\ and\ \citenamefont
  {Schollw\"ock}}]{Depenbrock2012a}%
  \BibitemOpen
  \bibfield  {author} {\bibinfo {author} {\bibfnamefont {S.}~\bibnamefont
  {Depenbrock}}, \bibinfo {author} {\bibfnamefont {I.~P.}\ \bibnamefont
  {McCulloch}}, \ and\ \bibinfo {author} {\bibfnamefont {U.}~\bibnamefont
  {Schollw\"ock}},\ }\href {\doibase 10.1103/PhysRevLett.109.067201} {\bibfield
   {journal} {\bibinfo  {journal} {Phys. Rev. Lett.}\ }\textbf {\bibinfo
  {volume} {109}},\ \bibinfo {pages} {067201} (\bibinfo {year}
  {2012})}\BibitemShut {NoStop}%
\bibitem [{\citenamefont {{Yan}}\ \emph {et~al.}(2011)\citenamefont {{Yan}},
  \citenamefont {{Huse}},\ and\ \citenamefont {{White}}}]{Yan_Huse2011a}%
  \BibitemOpen
  \bibfield  {author} {\bibinfo {author} {\bibfnamefont {S.}~\bibnamefont
  {{Yan}}}, \bibinfo {author} {\bibfnamefont {D.~A.}\ \bibnamefont {{Huse}}}, \
  and\ \bibinfo {author} {\bibfnamefont {S.~R.}\ \bibnamefont {{White}}},\
  }\href {\doibase 10.1126/science.1201080} {\bibfield  {journal} {\bibinfo
  {journal} {Science}\ }\textbf {\bibinfo {volume} {332}},\ \bibinfo {pages}
  {1173} (\bibinfo {year} {2011})},\ \Eprint {http://arxiv.org/abs/1011.6114}
  {arXiv:1011.6114 [cond-mat.str-el]} \BibitemShut {NoStop}%
\bibitem [{\citenamefont {{Ryu}}\ and\ \citenamefont
  {{Takayanagi}}(2006)}]{Ryu2006a}%
  \BibitemOpen
  \bibfield  {author} {\bibinfo {author} {\bibfnamefont {S.}~\bibnamefont
  {{Ryu}}}\ and\ \bibinfo {author} {\bibfnamefont {T.}~\bibnamefont
  {{Takayanagi}}},\ }\href {\doibase 10.1088/1126-6708/2006/08/045} {\bibfield
  {journal} {\bibinfo  {journal} {Journal of High Energy Physics}\ }\textbf
  {\bibinfo {volume} {8}},\ \bibinfo {eid} {045} (\bibinfo {year} {2006})},\
  \Eprint {http://arxiv.org/abs/hep-th/0605073} {hep-th/0605073} \BibitemShut
  {NoStop}%
\bibitem [{\citenamefont {Swingle}(2012)}]{Swingle2012a}%
  \BibitemOpen
  \bibfield  {author} {\bibinfo {author} {\bibfnamefont {B.}~\bibnamefont
  {Swingle}},\ }\href {\doibase 10.1103/PhysRevD.86.065007} {\bibfield
  {journal} {\bibinfo  {journal} {Phys. Rev. D}\ }\textbf {\bibinfo {volume}
  {86}},\ \bibinfo {pages} {065007} (\bibinfo {year} {2012})}\BibitemShut
  {NoStop}%
\bibitem [{\citenamefont {Zaletel}\ \emph {et~al.}(2013)\citenamefont
  {Zaletel}, \citenamefont {Mong},\ and\ \citenamefont
  {Pollmann}}]{Zaletel_2013a}%
  \BibitemOpen
  \bibfield  {author} {\bibinfo {author} {\bibfnamefont {M.~P.}\ \bibnamefont
  {Zaletel}}, \bibinfo {author} {\bibfnamefont {R.~S.~K.}\ \bibnamefont
  {Mong}}, \ and\ \bibinfo {author} {\bibfnamefont {F.}~\bibnamefont
  {Pollmann}},\ }\href {\doibase 10.1103/PhysRevLett.110.236801} {\bibfield
  {journal} {\bibinfo  {journal} {Phys. Rev. Lett.}\ }\textbf {\bibinfo
  {volume} {110}},\ \bibinfo {pages} {236801} (\bibinfo {year}
  {2013})}\BibitemShut {NoStop}%
\bibitem [{\citenamefont {{Liu}}\ \emph {et~al.}(2015)\citenamefont {{Liu}},
  \citenamefont {{Vaezi}}, \citenamefont {{Lee}},\ and\ \citenamefont
  {{Kim}}}]{Liu_Vaezi2015a}%
  \BibitemOpen
  \bibfield  {author} {\bibinfo {author} {\bibfnamefont {Z.}~\bibnamefont
  {{Liu}}}, \bibinfo {author} {\bibfnamefont {A.}~\bibnamefont {{Vaezi}}},
  \bibinfo {author} {\bibfnamefont {K.}~\bibnamefont {{Lee}}}, \ and\ \bibinfo
  {author} {\bibfnamefont {E.-A.}\ \bibnamefont {{Kim}}},\ }\href {\doibase
  10.1103/PhysRevB.92.081102} {\bibfield  {journal} {\bibinfo  {journal}
  {\prb}\ }\textbf {\bibinfo {volume} {92}},\ \bibinfo {eid} {081102} (\bibinfo
  {year} {2015})},\ \Eprint {http://arxiv.org/abs/1502.05391} {arXiv:1502.05391
  [cond-mat.str-el]} \BibitemShut {NoStop}%
\bibitem [{\citenamefont {{Wen}}(1990)}]{Wen1990a}%
  \BibitemOpen
  \bibfield  {author} {\bibinfo {author} {\bibfnamefont {X.~G.}\ \bibnamefont
  {{Wen}}},\ }\href {\doibase 10.1142/S0217979290000139} {\bibfield  {journal}
  {\bibinfo  {journal} {International Journal of Modern Physics B}\ }\textbf
  {\bibinfo {volume} {4}},\ \bibinfo {pages} {239} (\bibinfo {year}
  {1990})}\BibitemShut {NoStop}%
\bibitem [{\citenamefont {Nayak}\ \emph {et~al.}(2008)\citenamefont {Nayak},
  \citenamefont {Simon}, \citenamefont {Stern}, \citenamefont {Freedman},\ and\
  \citenamefont {Das~Sarma}}]{Nayak2008a}%
  \BibitemOpen
  \bibfield  {author} {\bibinfo {author} {\bibfnamefont {C.}~\bibnamefont
  {Nayak}}, \bibinfo {author} {\bibfnamefont {S.~H.}\ \bibnamefont {Simon}},
  \bibinfo {author} {\bibfnamefont {A.}~\bibnamefont {Stern}}, \bibinfo
  {author} {\bibfnamefont {M.}~\bibnamefont {Freedman}}, \ and\ \bibinfo
  {author} {\bibfnamefont {S.}~\bibnamefont {Das~Sarma}},\ }\href {\doibase
  10.1103/RevModPhys.80.1083} {\bibfield  {journal} {\bibinfo  {journal} {Rev.
  Mod. Phys.}\ }\textbf {\bibinfo {volume} {80}},\ \bibinfo {pages} {1083}
  (\bibinfo {year} {2008})}\BibitemShut {NoStop}%
\bibitem [{Com({\natexlab{a}})}]{Comment4}%
  \BibitemOpen
  \href@noop {} {\bibfield  {journal} {\bibinfo  {journal} {It must be noted that $\epsilon_{2}^{\para{\chi_a}}\para{a,b}$ is not a metric in the mathematical sense, e.g., $\epsilon_{2}^{\para{\chi_a}}\para{a,b} \neq \epsilon_{2}^{\para{\chi_b}}\para{b,a}$. However, it can serve as a good physical measure to gauge the distance between two quantum states.}\ }
  }\BibitemShut {NoStop}%
\bibitem [{\citenamefont {{Alcaraz}}\ \emph {et~al.}(2011)\citenamefont
  {{Alcaraz}}, \citenamefont {{Berganza}},\ and\ \citenamefont
  {{Sierra}}}]{Alcaraz2011a}%
  \BibitemOpen
  \bibfield  {author} {\bibinfo {author} {\bibfnamefont {F.~C.}\ \bibnamefont
  {{Alcaraz}}}, \bibinfo {author} {\bibfnamefont {M.~I.}\ \bibnamefont
  {{Berganza}}}, \ and\ \bibinfo {author} {\bibfnamefont {G.}~\bibnamefont
  {{Sierra}}},\ }\href {\doibase 10.1103/PhysRevLett.106.201601} {\bibfield
  {journal} {\bibinfo  {journal} {Physical Review Letters}\ }\textbf {\bibinfo
  {volume} {106}},\ \bibinfo {eid} {201601} (\bibinfo {year} {2011})},\ \Eprint
  {http://arxiv.org/abs/1101.2881} {arXiv:1101.2881 [cond-mat.stat-mech]}
  \BibitemShut {NoStop}%
\bibitem [{\citenamefont {{Ib{\'a}{\~n}ez Berganza}}\ \emph
  {et~al.}(2012)\citenamefont {{Ib{\'a}{\~n}ez Berganza}}, \citenamefont
  {{Castilho Alcaraz}},\ and\ \citenamefont {{Sierra}}}]{Berganza2011a}%
  \BibitemOpen
  \bibfield  {author} {\bibinfo {author} {\bibfnamefont {M.}~\bibnamefont
  {{Ib{\'a}{\~n}ez Berganza}}}, \bibinfo {author} {\bibfnamefont
  {F.}~\bibnamefont {{Castilho Alcaraz}}}, \ and\ \bibinfo {author}
  {\bibfnamefont {G.}~\bibnamefont {{Sierra}}},\ }\href {\doibase
  10.1088/1742-5468/2012/01/P01016} {\bibfield  {journal} {\bibinfo  {journal}
  {Journal of Statistical Mechanics: Theory and Experiment}\ }\textbf {\bibinfo
  {volume} {1}},\ \bibinfo {pages} {01016} (\bibinfo {year} {2012})},\ \Eprint
  {http://arxiv.org/abs/1109.5673} {arXiv:1109.5673 [cond-mat.stat-mech]}
  \BibitemShut {NoStop}%
\bibitem [{\citenamefont {Taddia}\ \emph {et~al.}(2016)\citenamefont {Taddia},
  \citenamefont {Ortolani},\ and\ \citenamefont {Pálmai}}]{Taddia2016a}%
  \BibitemOpen
  \bibfield  {author} {\bibinfo {author} {\bibfnamefont {L.}~\bibnamefont
  {Taddia}}, \bibinfo {author} {\bibfnamefont {F.}~\bibnamefont {Ortolani}}, \
  and\ \bibinfo {author} {\bibfnamefont {T.}~\bibnamefont {Pálmai}},\ }\href
  {http://stacks.iop.org/1742-5468/2016/i=9/a=093104} {\bibfield  {journal}
  {\bibinfo  {journal} {Journal of Statistical Mechanics: Theory and
  Experiment}\ }\textbf {\bibinfo {volume} {2016}},\ \bibinfo {pages} {093104}
  (\bibinfo {year} {2016})}\BibitemShut {NoStop}%
\bibitem [{\citenamefont {{Astaneh}}\ and\ \citenamefont
  {{Mosaffa}}(2013)}]{Astaneh2013a}%
  \BibitemOpen
  \bibfield  {author} {\bibinfo {author} {\bibfnamefont {A.~F.}\ \bibnamefont
  {{Astaneh}}}\ and\ \bibinfo {author} {\bibfnamefont {A.~E.}\ \bibnamefont
  {{Mosaffa}}},\ }\href {\doibase 10.1007/JHEP03(2013)135} {\bibfield
  {journal} {\bibinfo  {journal} {Journal of High Energy Physics}\ }\textbf
  {\bibinfo {volume} {3}},\ \bibinfo {eid} {135} (\bibinfo {year}
  {2013})}\BibitemShut {NoStop}%
\bibitem [{\citenamefont {{Fradkin}}\ and\ \citenamefont
  {{Kadanoff}}(1980)}]{Fradkin_kadanoff1980a}%
  \BibitemOpen
  \bibfield  {author} {\bibinfo {author} {\bibfnamefont {E.}~\bibnamefont
  {{Fradkin}}}\ and\ \bibinfo {author} {\bibfnamefont {L.~P.}\ \bibnamefont
  {{Kadanoff}}},\ }\href {\doibase 10.1016/0550-3213(80)90472-1} {\bibfield
  {journal} {\bibinfo  {journal} {Nuclear Physics B}\ }\textbf {\bibinfo
  {volume} {170}},\ \bibinfo {pages} {1} (\bibinfo {year} {1980})}\BibitemShut
  {NoStop}%
\bibitem [{\citenamefont {{Fendley}}(2012)}]{Fendley2012a}%
  \BibitemOpen
  \bibfield  {author} {\bibinfo {author} {\bibfnamefont {P.}~\bibnamefont
  {{Fendley}}},\ }\href {\doibase 10.1088/1742-5468/2012/11/P11020} {\bibfield
  {journal} {\bibinfo  {journal} {Journal of Statistical Mechanics: Theory and
  Experiment}\ }\textbf {\bibinfo {volume} {11}},\ \bibinfo {pages} {11020}
  (\bibinfo {year} {2012})},\ \Eprint {http://arxiv.org/abs/1209.0472}
  {arXiv:1209.0472 [cond-mat.str-el]} \BibitemShut {NoStop}%
\bibitem [{\citenamefont {{Ortiz}}\ \emph {et~al.}(2012)\citenamefont
  {{Ortiz}}, \citenamefont {{Cobanera}},\ and\ \citenamefont
  {{Nussinov}}}]{Ortiz_Cobanera2011a}%
  \BibitemOpen
  \bibfield  {author} {\bibinfo {author} {\bibfnamefont {G.}~\bibnamefont
  {{Ortiz}}}, \bibinfo {author} {\bibfnamefont {E.}~\bibnamefont {{Cobanera}}},
  \ and\ \bibinfo {author} {\bibfnamefont {Z.}~\bibnamefont {{Nussinov}}},\
  }\href {\doibase 10.1016/j.nuclphysb.2011.09.012} {\bibfield  {journal}
  {\bibinfo  {journal} {Nuclear Physics B}\ }\textbf {\bibinfo {volume}
  {854}},\ \bibinfo {pages} {780} (\bibinfo {year} {2012})},\ \Eprint
  {http://arxiv.org/abs/1108.2276} {arXiv:1108.2276 [cond-mat.stat-mech]}
  \BibitemShut {NoStop}%
\bibitem [{\citenamefont {{Vaezi}}\ and\ \citenamefont
  {{Kim}}(2013)}]{Vaezi_Kim2013a}%
  \BibitemOpen
  \bibfield  {author} {\bibinfo {author} {\bibfnamefont {A.}~\bibnamefont
  {{Vaezi}}}\ and\ \bibinfo {author} {\bibfnamefont {E.-A.}\ \bibnamefont
  {{Kim}}},\ }\href@noop {} {\bibfield  {journal} {\bibinfo  {journal} {ArXiv
  e-prints}\ } (\bibinfo {year} {2013})},\ \Eprint
  {http://arxiv.org/abs/1310.7434} {arXiv:1310.7434 [cond-mat.str-el]}
  \BibitemShut {NoStop}%
\bibitem [{Com({\natexlab{b}})}]{Comment0}%
  \BibitemOpen
  \href@noop {} {\bibfield  {journal} {\bibinfo  {journal} {This can be
  understood through the following argument for the $Z_2$ parafermion chain
  (Kitaev's 1D model) at $t1\gg t2$ limit which can be trivially generalized to
  other $Z_N$ models. In this limit the two-fold degenerate GSs are :
  $\ket{gs}_m = \frac{1}{\sqrt{2}}\sum_{n=0,1}(-1)^{mn}\ket{n \cdots n}_L
  \ket{n\cdots n}_R$ ($m=0,1$) which have well-defined fermion parity $P =
  \prod_j \sigma_x(i) = (-1)^m$. Since for both states $\rho_L(m) = 1/2 \para{
  \ket{0 \cdots 0}_L \bra{0 \cdots 0}_L + \ket{1 \cdots 1}_L \bra{1 \cdots
  1}_L} $ the entanglement distance is zero}\ } }\BibitemShut
  {NoStop}%
\bibitem [{\citenamefont {Chung}\ and\ \citenamefont
  {Peschel}(2001)}]{Chung2001a}%
  \BibitemOpen
  \bibfield  {author} {\bibinfo {author} {\bibfnamefont {M.-C.}\ \bibnamefont
  {Chung}}\ and\ \bibinfo {author} {\bibfnamefont {I.}~\bibnamefont
  {Peschel}},\ }\href {\doibase 10.1103/PhysRevB.64.064412} {\bibfield
  {journal} {\bibinfo  {journal} {Phys. Rev. B}\ }\textbf {\bibinfo {volume}
  {64}},\ \bibinfo {pages} {064412} (\bibinfo {year} {2001})}\BibitemShut
  {NoStop}%
\bibitem [{\citenamefont {Cheong}\ and\ \citenamefont
  {Henley}(2004)}]{Cheong2004a}%
  \BibitemOpen
  \bibfield  {author} {\bibinfo {author} {\bibfnamefont {S.-A.}\ \bibnamefont
  {Cheong}}\ and\ \bibinfo {author} {\bibfnamefont {C.~L.}\ \bibnamefont
  {Henley}},\ }\href {\doibase 10.1103/PhysRevB.69.075111} {\bibfield
  {journal} {\bibinfo  {journal} {Phys. Rev. B}\ }\textbf {\bibinfo {volume}
  {69}},\ \bibinfo {pages} {075111} (\bibinfo {year} {2004})}\BibitemShut
  {NoStop}%
\bibitem [{\citenamefont {Qi}\ \emph {et~al.}(2012)\citenamefont {Qi},
  \citenamefont {Katsura},\ and\ \citenamefont {Ludwig}}]{Qi_Katsura2012a}%
  \BibitemOpen
  \bibfield  {author} {\bibinfo {author} {\bibfnamefont {X.-L.}\ \bibnamefont
  {Qi}}, \bibinfo {author} {\bibfnamefont {H.}~\bibnamefont {Katsura}}, \ and\
  \bibinfo {author} {\bibfnamefont {A.~W.~W.}\ \bibnamefont {Ludwig}},\ }\href
  {\doibase 10.1103/PhysRevLett.108.196402} {\bibfield  {journal} {\bibinfo
  {journal} {Phys. Rev. Lett.}\ }\textbf {\bibinfo {volume} {108}},\ \bibinfo
  {pages} {196402} (\bibinfo {year} {2012})}\BibitemShut {NoStop}%
\bibitem [{\citenamefont {Laughlin}(1983)}]{Laughlin1983a}%
  \BibitemOpen
  \bibfield  {author} {\bibinfo {author} {\bibfnamefont {R.~B.}\ \bibnamefont
  {Laughlin}},\ }\href {\doibase 10.1103/PhysRevLett.50.1395} {\bibfield
  {journal} {\bibinfo  {journal} {Phys. Rev. Lett.}\ }\textbf {\bibinfo
  {volume} {50}},\ \bibinfo {pages} {1395} (\bibinfo {year}
  {1983})}\BibitemShut {NoStop}%
\bibitem [{\citenamefont {{Clarke}}\ \emph {et~al.}(2013)\citenamefont
  {{Clarke}}, \citenamefont {{Alicea}},\ and\ \citenamefont
  {{Shtengel}}}]{Clarke2012a}%
  \BibitemOpen
  \bibfield  {author} {\bibinfo {author} {\bibfnamefont {D.~J.}\ \bibnamefont
  {{Clarke}}}, \bibinfo {author} {\bibfnamefont {J.}~\bibnamefont {{Alicea}}},
  \ and\ \bibinfo {author} {\bibfnamefont {K.}~\bibnamefont {{Shtengel}}},\
  }\href {\doibase 10.1038/ncomms2340} {\bibfield  {journal} {\bibinfo
  {journal} {Nature Communications}\ }\textbf {\bibinfo {volume} {4}},\
  \bibinfo {eid} {1348} (\bibinfo {year} {2013})},\ \Eprint
  {http://arxiv.org/abs/1204.5479} {arXiv:1204.5479 [cond-mat.str-el]}
  \BibitemShut {NoStop}%
\bibitem [{\citenamefont {Lindner}\ \emph {et~al.}(2012)\citenamefont
  {Lindner}, \citenamefont {Berg}, \citenamefont {Refael},\ and\ \citenamefont
  {Stern}}]{Lindner_Berg2013a}%
  \BibitemOpen
  \bibfield  {author} {\bibinfo {author} {\bibfnamefont {N.~H.}\ \bibnamefont
  {Lindner}}, \bibinfo {author} {\bibfnamefont {E.}~\bibnamefont {Berg}},
  \bibinfo {author} {\bibfnamefont {G.}~\bibnamefont {Refael}}, \ and\ \bibinfo
  {author} {\bibfnamefont {A.}~\bibnamefont {Stern}},\ }\href {\doibase
  10.1103/PhysRevX.2.041002} {\bibfield  {journal} {\bibinfo  {journal} {Phys.
  Rev. X}\ }\textbf {\bibinfo {volume} {2}},\ \bibinfo {pages} {041002}
  (\bibinfo {year} {2012})}\BibitemShut {NoStop}%
\bibitem [{\citenamefont {Cheng}(2012)}]{Cheng2012a}%
  \BibitemOpen
  \bibfield  {author} {\bibinfo {author} {\bibfnamefont {M.}~\bibnamefont
  {Cheng}},\ }\href {\doibase 10.1103/PhysRevB.86.195126} {\bibfield  {journal}
  {\bibinfo  {journal} {Phys. Rev. B}\ }\textbf {\bibinfo {volume} {86}},\
  \bibinfo {pages} {195126} (\bibinfo {year} {2012})}\BibitemShut {NoStop}%
\bibitem [{\citenamefont {Vaezi}(2013)}]{Vaezi_ftsc_2013a}%
  \BibitemOpen
  \bibfield  {author} {\bibinfo {author} {\bibfnamefont {A.}~\bibnamefont
  {Vaezi}},\ }\href {\doibase 10.1103/PhysRevB.87.035132} {\bibfield  {journal}
  {\bibinfo  {journal} {Phys. Rev. B}\ }\textbf {\bibinfo {volume} {87}},\
  \bibinfo {pages} {035132} (\bibinfo {year} {2013})}\BibitemShut {NoStop}%
\bibitem [{\citenamefont {Barkeshli}\ \emph {et~al.}(2013)\citenamefont
  {Barkeshli}, \citenamefont {Jian},\ and\ \citenamefont
  {Qi}}]{Barkeshli_Jian2013a}%
  \BibitemOpen
  \bibfield  {author} {\bibinfo {author} {\bibfnamefont {M.}~\bibnamefont
  {Barkeshli}}, \bibinfo {author} {\bibfnamefont {C.-M.}\ \bibnamefont {Jian}},
  \ and\ \bibinfo {author} {\bibfnamefont {X.-L.}\ \bibnamefont {Qi}},\ }\href
  {\doibase 10.1103/PhysRevB.87.045130} {\bibfield  {journal} {\bibinfo
  {journal} {Phys. Rev. B}\ }\textbf {\bibinfo {volume} {87}},\ \bibinfo
  {pages} {045130} (\bibinfo {year} {2013})}\BibitemShut {NoStop}%
\bibitem [{\citenamefont {You}\ and\ \citenamefont {Wen}(2012)}]{You_Wen2012a}%
  \BibitemOpen
  \bibfield  {author} {\bibinfo {author} {\bibfnamefont {Y.-Z.}\ \bibnamefont
  {You}}\ and\ \bibinfo {author} {\bibfnamefont {X.-G.}\ \bibnamefont {Wen}},\
  }\href {\doibase 10.1103/PhysRevB.86.161107} {\bibfield  {journal} {\bibinfo
  {journal} {Phys. Rev. B}\ }\textbf {\bibinfo {volume} {86}},\ \bibinfo
  {pages} {161107} (\bibinfo {year} {2012})}\BibitemShut {NoStop}%
\bibitem [{\citenamefont {{Dong}}\ \emph {et~al.}(2008)\citenamefont {{Dong}},
  \citenamefont {{Fradkin}}, \citenamefont {{Leigh}},\ and\ \citenamefont
  {{Nowling}}}]{Dong_Fradkin2008a}%
  \BibitemOpen
  \bibfield  {author} {\bibinfo {author} {\bibfnamefont {S.}~\bibnamefont
  {{Dong}}}, \bibinfo {author} {\bibfnamefont {E.}~\bibnamefont {{Fradkin}}},
  \bibinfo {author} {\bibfnamefont {R.~G.}\ \bibnamefont {{Leigh}}}, \ and\
  \bibinfo {author} {\bibfnamefont {S.}~\bibnamefont {{Nowling}}},\ }\href
  {\doibase 10.1088/1126-6708/2008/05/016} {\bibfield  {journal} {\bibinfo
  {journal} {Journal of High Energy Physics}\ }\textbf {\bibinfo {volume}
  {5}},\ \bibinfo {eid} {016} (\bibinfo {year} {2008})},\ \Eprint
  {http://arxiv.org/abs/0802.3231} {arXiv:0802.3231 [hep-th]} \BibitemShut
  {NoStop}%
  \bibitem [{\citenamefont {Haldane}(1983)}]{Haldane1983a}%
  \BibitemOpen
  \bibfield  {author} {\bibinfo {author} {\bibfnamefont {F.~D.~M.}\
  \bibnamefont {Haldane}},\ }\href {\doibase 10.1103/PhysRevLett.51.605}
  {\bibfield  {journal} {\bibinfo  {journal} {Phys. Rev. Lett.}\ }\textbf
  {\bibinfo {volume} {51}},\ \bibinfo {pages} {605} (\bibinfo {year}
  {1983})}\BibitemShut {NoStop}%
\bibitem [{\citenamefont {Wilson}(1975)}]{Wilson1975a}%
  \BibitemOpen
  \bibfield  {author} {\bibinfo {author} {\bibfnamefont {K.~G.}\ \bibnamefont
  {Wilson}},\ }\href {\doibase 10.1103/RevModPhys.47.773} {\bibfield  {journal}
  {\bibinfo  {journal} {Rev. Mod. Phys.}\ }\textbf {\bibinfo {volume} {47}},\
  \bibinfo {pages} {773} (\bibinfo {year} {1975})}\BibitemShut {NoStop}%
\bibitem [{\citenamefont {{S{\'a}rosi}}\ and\ \citenamefont
  {{Ugajin}}(2016)}]{Ugajin2016a}%
  \BibitemOpen
  \bibfield  {author} {\bibinfo {author} {\bibfnamefont {G.}~\bibnamefont
  {{S{\'a}rosi}}}\ and\ \bibinfo {author} {\bibfnamefont {T.}~\bibnamefont
  {{Ugajin}}},\ }\href {\doibase 10.1007/JHEP07(2016)114} {\bibfield  {journal}
  {\bibinfo  {journal} {Journal of High Energy Physics}\ }\textbf {\bibinfo
  {volume} {7}},\ \bibinfo {eid} {114} (\bibinfo {year} {2016})}\BibitemShut
  {NoStop}%
\end{thebibliography}
%
\end{document}